\documentclass[aps,pra,floatfix,reprint]{revtex4-2} 
\usepackage{graphicx}
\usepackage{natbib}
\usepackage{braket}
\usepackage{amsmath} 
\usepackage{amssymb}
\usepackage{bm}
\usepackage[separate-uncertainty=true, multi-part-units = single, product-units=single]{siunitx}
\usepackage{upgreek}
\usepackage{color}
\usepackage{dcolumn}
\usepackage{soul} 
\newcolumntype{d}[1]{D..{#1}}

\begin{document}

\title{Cryogenic ion trap system for high-fidelity near-field microwave-driven quantum logic}

\author{M.\,A.\,Weber}
\author{C.\,L\"oschnauer} 
\author{J.\,Wolf}
\author{M.\,F.\,Gely} 
\author{R.\,K.\,Hanley} 
\author{J.\,F.\,Goodwin} 
\author{C.\,J.\,Ballance} 
\author{T.\,P.\,Harty} 
\author{D.\,M.\,Lucas} 
\affiliation{Clarendon Laboratory, Department of Physics, University of Oxford, Parks Road, Oxford OX1 3PU, U.K.}

\date{22 July 2022} 

\begin{abstract}
We report the design, fabrication, and characterization of a cryogenic ion trap system for the implementation of quantum logic driven by near-field microwaves.
The trap incorporates an on-chip microwave resonator with an electrode geometry designed to null the microwave field component that couples directly to the qubit, while giving a large field gradient for driving entangling logic gates.
We map the microwave field using a single $^{43}$Ca$^+$ ion, and measure the ion trapping lifetime and motional mode heating rates for one and two ions.

\end{abstract}

\maketitle
Trapped ions are a promising candidate for building a general-purpose quantum processor, with both single-qubit \cite{Harty2014} and two-qubit \cite{Gaebler2016, Ballance2016, Srinivas2021, Clark2021} gates achieved at the demanding fidelities required for quantum error correction \cite{Gottesman1998, Raussendorf2007}.
Trapped-ion qubits are typically implemented using electric-dipole-forbidden transitions where the state lifetime is sufficiently large that decoherence via spontaneous emission is negligible, leading to coherence times of the order of minutes \cite{Langer2005, Haffner2005, Sepiol2019} or longer \cite{Wang2017}.
The qubit transitions typically lie in the optical domain operating on electric-quadrupole transitions \cite{Nagerl2000}, or in the microwave domain between hyperfine states within the same manifold \cite{Monroe1995}.
Although hyperfine qubits lie in the microwave domain, they are usually manipulated using stimulated Raman transitions with tightly focused laser beams, as the short optical wavelength enables single-qubit addressing \cite{Nagerl1999} and efficient coupling between the ions' spin and motional degrees of freedom \cite{Wineland1998}.
Laser-driven operations utilising stimulated Raman transitions are fundamentally limited to an infidelity of $\sim 10^{-4}$ \cite{Ozeri2007, Schafer2018} arising from photon scattering.
Furthermore, the scaling of stimulated Raman operations to a large-scale quantum processor is challenging as many high-intensity laser beams need to be controlled and aligned with sub-$\upmu$m precision.

Microwave radiation can be used to drive hyperfine or Zeeman qubits directly.
However, as the free-space wavelength of microwave radiation is much greater than the spatial extent of the trapped ions' wave-packet, the resulting spin-motion coupling is weak, making multi-qubit operations impractical.
The spin-motion coupling can be increased by several orders of magnitude if one is able to engineer a large spatial gradient in the microwave field.
One way to achieve an effective microwave field gradient is by combining far-field microwaves with a strong, static magnetic field gradient \cite{Mintert2001, Johanning2009, Lake2015}.
However, this method requires radiative atomic-dressing techniques \cite{Webster2013, Mikelsons2015, Cohen2015} to minimise decoherence because the qubit states need to be first-order sensitive to magnetic fields.
An alternative solution is to position the ions in the near-field regime of a current-carrying conductor \cite{Wineland1998, Ospelkaus2008, Ospelkaus2011}; here the field gradient is determined by the distance to the conductor and conductor geometry rather than the free-space wavelength of the microwaves.
In addition to these methods, a novel spin-motion coupling has been recently demonstrated using a radio-frequency field gradient oscillating close to the ions' motional frequency \cite{Srinivas2019}.

Microwave technology is much more mature than laser technology, and is used in many everyday devices such as mobile telephony.
It is commercially available at a considerably lower cost than laser systems, and is also significantly easier to control.
Microwave circuitry can also be directly integrated into control structures, which facilitates the production of chip-based ion traps that are amenable to scaling into quantum `CCD-like' devices \cite{Wineland1998, Kielpinski2002, McHugh2005, Chiaverini2008, Monroe2013}.
There have been significant advances in recent years using surface traps with single-layer \cite{Ospelkaus2011, Allcock2013, Carsjens2013} and multi-layer \cite{Bautista2019, Hahn2019} integrated microwave circuit elements to perform quantum logic gates, with the highest-fidelity two-qubit operations \cite{Harty2016, Zarantonello2019, Srinivas2021} approaching the fidelities of state-of-the-art laser-based systems \cite{Gaebler2016, Ballance2016, Clark2021}.

This paper details the design, construction, and characterisation of a surface-electrode trap with integrated microwave circuitry, which is designed for implementing microwave-driven two-qubit gates with significantly improved fidelity and/or speed compared with that achieved in our first-generation system ($F=0.997$ in $t_g=\SI{3.25}{ms}$) \cite{Harty2016}.
The apparatus features a chip trap with a novel microwave electrode concept, designed for a previously unused hyperfine ``atomic clock'' qubit \cite{Langer2005,Sepiol2019} in $^{43}\rm{Ca}^+$, operating at a static magnetic field of $B_0 \approx \SI{288}{G}$.
The surface trap is designed to operate at cryogenic temperatures to mitigate gate errors due to motional mode heating, to reduce ion loss from background gas collisions, and to increase the achievable microwave field gradient.
For the purpose of characterizing performance, the system is designed to operate at any temperature from cryogenic ($\sim$\SI{20}{K}) to ambient (\SI{300}{K}).

The remainder of this paper is structured as follows.
In section~\ref{sec:improvements} we outline the desired improvements relative to our first-generation experiment \cite{Harty2016}, before describing the design and fabrication of the surface trap in sections~\ref{sec:trapdesign} and~\ref{sec:trapfab}.
We detail in sections~\ref{sec:filterboards} and~\ref{sec:cryo} the construction of the experimental apparatus which enables the surface trap to be operated at cryogenic temperature $T\sim\SI{20}{K}$.
In section~\ref{sec:results} we present characterisation measurements of the surface trap and discuss the future prospects for high-speed and high-fidelity microwave-driven quantum logic in this system.
        
\section{Targeted Improvements}
\label{sec:improvements}

The experiment targets improvements to the two-qubit gate speed and fidelity by employing an improved trap design and choice of qubit.
A particular focus is placed on gate speed improvements as several error sources for entangling gates increase with the gate duration \cite{Ballance2016}.

In the near-field microwave regime, the spin-motion coupling strength is proportional to the spatial gradient of the microwave magnetic field.
Therefore, the two-qubit gate speed can be improved by increasing this magnetic field gradient.
Naively, this can be achieved by increasing the injected microwave power $P_{\upmu{\rm{W}}}$.
However, a power increase is not a practical solution as the field gradient scales as $\sqrt{P_{\upmu{\rm{W}}}}$, leading to an unrealistic input power and trap power-handling requirements for a modest increase in gate speed.
A more efficient approach is to reduce the ion-to-electrode distance $d$, as the field gradient scales as $\sim 1/d^2$.
A disadvantage of the reduced ion-to-electrode distance is the likely increase in electric field noise, which has been observed to scale approximately as $\sim 1/d^4$ \cite{Sedlacek2018}.
Greater electric field noise increases two-qubit gate errors and therefore needs to be balanced against any gain from an increased gate speed.
To mitigate the field noise, one can cool the trap electrodes to cryogenic temperatures; this technique has been shown to reduce electric field noise by several orders of magnitude \cite{Labaziewicz2008, Deslauriers2006}.

The two-qubit gate fidelity can also be improved by increasing the ratio of microwave field gradient to field amplitude.
The field gradient drives the two-qubit gate, whilst the field amplitude leads to unwanted AC Zeeman shifts, and can drive undesired magnetic dipole transitions.
One could interfere fields from multiple independent microwave electrodes to produce a large field gradient with a perfect microwave-null \cite{Ospelkaus2008}.
However, in practice the active stabilisation of all microwave current amplitudes and phases with sufficient accuracy in multiple electrodes is a significant challenge.
An alternative approach is to design a trap which creates a microwave null passively \cite{Carsjens2013}, eliminating technical differential amplitude and phase noise.

As well as the trap design considerations above, we aim to improve the gate speed and fidelity through a more favourable choice of qubit states than in our previous work \cite{Harty2016}.
In that experiment, we employed a $\left|\Delta M\right| = 1$ qubit transition.
Such a transition can be driven by a linearly-polarised microwave field oriented orthogonal to the quantisation axis.
This field creates an equal superposition of left- and right-circularly polarised fields, hence only half the applied microwave power is available to drive the qubit transition.
It is therefore beneficial to utilise a $\Delta M = 0$ qubit transition where all of the available microwave field gradient can be used to drive the spin-motion coupling.

%

\begin{figure}
\includegraphics[width=\columnwidth]{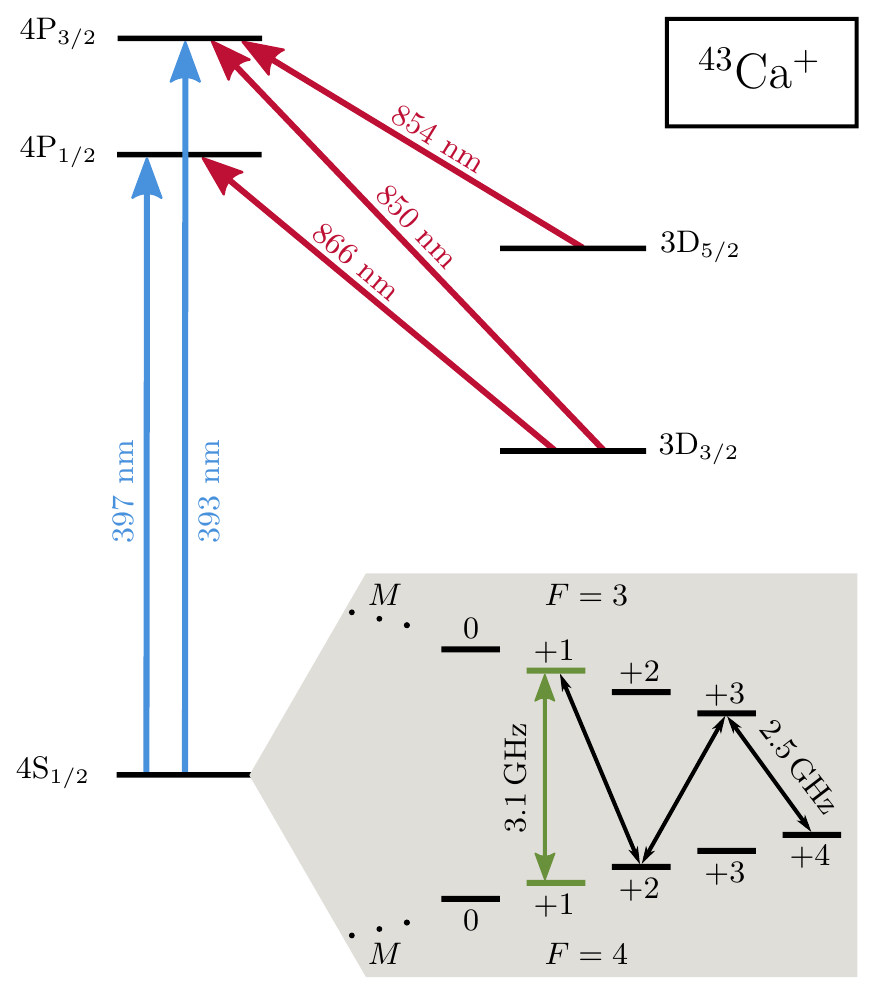}
\caption{\label{fig:energy_levels}Energy level diagram of $^{43}{\rm{Ca}}^+$ showing the relevant transitions for laser cooling, state preparation, and readout.
The lower panel shows relevant states within the hyperfine structure of the 4S$_{1/2}$ ground level, at a static magnetic field of $B_0 = \SI{287.783}{G}$.
The green states show the clock transition used for the qubit in this work, and the black arrows show the transitions used for state preparation and readout.
Zeeman splittings between adjacent $M$ states are $\approx \SI{110}{MHz}$.}
\end{figure}

We previously used the ``clock'' qubit states $\left|F=4, M=0\right\rangle$ and $\left|F=3, M=+1\right\rangle$ in the 4S$_{1/2}$ ground level hyperfine manifold of $^{43}{\rm{Ca}}^+$ (figure~\ref{fig:energy_levels}), operating at a static field of $B_0\approx\SI{146}{G}$, where the $\left|\Delta M\right| = 1$ qubit transition energy is first-order insensitive to magnetic field fluctuations \cite{Harty2014}.
In this second-generation experiment, we use a $\Delta M = 0$ clock qubit operating on the $\left|F=4, M=+1\right\rangle \leftrightarrow \left|F=3, M=+1\right\rangle$ transition at $B_0 = \SI{287.783}{G}$, allowing efficient coupling of the available microwave gradient to drive the entangling operation.
In addition to the efficient microwave gradient polarisation, the intrinsic transition strength of the $\Delta M = 0$ qubit is $\approx \sqrt{2}$ larger than that of the $\left|\Delta M\right| = 1$ qubit.
Together, these factors increase the achievable two-qubit operation speed by a factor of $\approx 2$ for a given microwave power.
Alternatively, when operating at the same gate speed, AC Zeeman shifts and driving of undesired spectator transitions are reduced.
The increase in the static magnetic field from \SI{146}{G} to \SI{288}{G} approximately doubles the detuning of spectator transitions from the qubit transition, further suppressing their effects.

In addition to the change of qubit, the ion-to-electrode distance in this new surface trap is $d = \SI{40}{\micro m}$, approximately half that of the previous trap.
This increases the achievable microwave field gradient and hence the two-qubit gate speed by an additional factor $\sim 4$ and is complemented by an electrode layout which gives a passive microwave null.
To mitigate the anticipated increase in the motional mode heating rate $\dot{\bar{n}}$ due to smaller $d$, the surface trap is designed to operate at cryogenic temperatures.
An added benefit of cryogenic operation is the reduction in resistive losses within the microwave electrode, which leads to higher achievable microwave currents and thus field gradients.
It also improves the vacuum, giving fewer background gas collisions and longer ion trapping lifetime.

\section{Surface trap design}
\label{sec:trapdesign}
    
\begin{figure}
\includegraphics[width=\columnwidth]{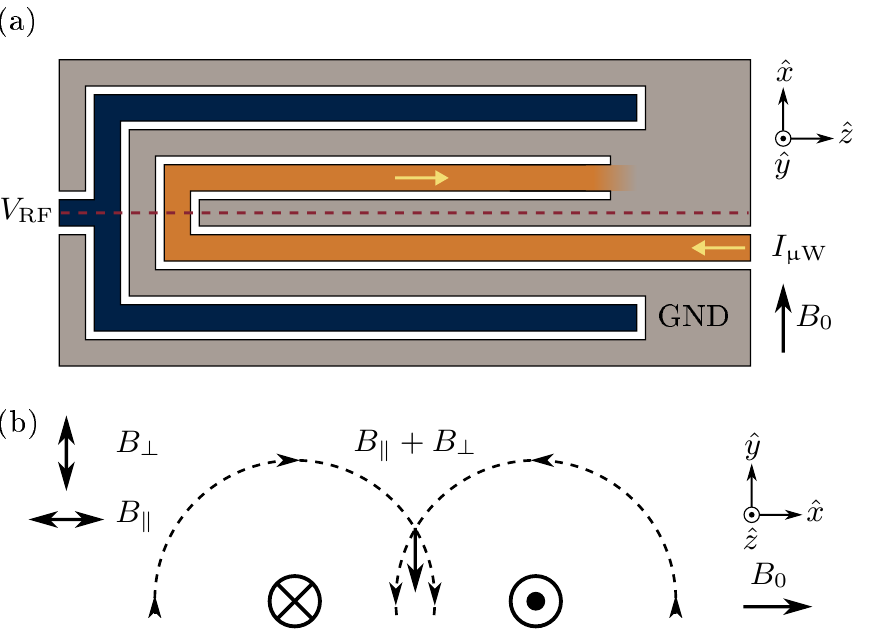}
\caption{%
(a) Simplified top view of the surface trap (not to scale), illustrating the operational principles.
The blue, orange, and grey regions correspond to the RF, microwave, and ground plane electrodes respectively.
The red dashed line shows the position of the RF null, and $B_0$ is the applied static magnetic field.
(b) A cross-sectional view of the instantaneous magnetic field created by current flowing in the microwave electrodes (shown as one-dimensional wires for illustration).}
\label{fig:trap_schematic}
\end{figure}

The use of symmetry to passively null the field amplitude eliminates the need for active stabilisation of multiple microwave signals, reducing potential sources of gate error.
A `meander' electrode structure has previously been shown to passively null the microwave field in all three spatial axes \cite{Carsjens2013, Wahnschaffe2017}.
The position of the minimum relies on a complex interference pattern between multiple sections of the electrode, making the surface trap design sensitive to the geometry and microwave current distribution.
This is particularly problematic for designs relying on a single metal layer, where the ground plane is effectively bisected by the meander and RF electrodes.
The design of a meander trap that aims to minimise the field amplitude but maximise the field gradient is therefore difficult as the optimisation is multi-dimensional and not constrained by symmetries, leading to difficulties finding a broad global minimum.
Furthermore, as the meander passively nulls the microwave field along all axes, it has the undesirable side-effect of restricting the achievable Rabi frequency used for state preparation.
An additional electrode can be added outside the meander \cite{Wahnschaffe2017}; however this is heavily screened by the ground plane, resulting in a weak field at the trap centre and hence slow state preparation.

To address the challenges associated with the meander design, we have designed a surface trap using a single, resonantly-coupled `U-shaped' microwave electrode (see figure \ref{fig:trap_schematic}(a)) which is terminated on the surface trap ground plane to form a quarter-wave ($\lambda/4$) resonator \cite{Pozar2011}.
The `U-shaped' microwave electrode design only results in a passive microwave field null along a single axis, thereby allowing for faster state preparation.
Further, the resonant structure allows for efficient coupling of microwave currents onto the surface trap.
A $\lambda/4$ resonator was chosen over a $\lambda/2$ resonator as the shorter microwave electrode length limits resistive losses.
Additionally, a $\lambda/4$ resonator allows the microwave electrode to be grounded, hence avoiding photoelectric charging effects.
The inherent symmetry of the `U-shaped' microwave electrode reduces the number of free parameters in the optimisation of a microwave minimum, making global optimisation more feasible.
The symmetry also increases the robustness of the trap to manufacturing imperfections, such as a global reduction in electrode width, as only asymmetric imperfections break the symmetry.

The basic design concept is illustrated in figure \ref{fig:trap_schematic}(b), where we consider a cross-sectional view of the current in the microwave electrode at a single instant in time.
Along the symmetry axis the magnetic field components parallel to $\hat{x}$ destructively interfere, resulting in a field null.
As one moves closer to either section of the electrode, the field component $B_x$ increases.
With the static magnetic field parallel to $\hat{x}$, the required $\pi$-polarised field gradient is created, while the $\pi$-field that couples directly to the qubit is nulled.
In reality, a small phase difference in the two opposing currents results in a minimum instead of a perfect null (see Sec.~\ref{sec:MW_minimum_simplistic}).
Along $\hat{y}$ the fields constructively interfere, resulting in a non-zero field component $B_y$ which facilitates a large Rabi frequency to drive both $\sigma^+$ and $\sigma^-$ transitions for state preparation in the hyperfine manifold without the need for additional electrodes.
The presence of such a large $\sigma$-field component which does not couple to the qubit may appear detrimental to the entangling operation; however we have shown in our previous work that this is not the limiting source of infidelity \cite{Harty2016}.

The microwave electrode was designed to be a resonant structure with a modest $Q$-factor ($Q_\text{tot} \approx 10$), with a centre frequency of $\SI{3.1}{GHz}$ to match the hyperfine splitting of the $\left|F=4,M=+1\right\rangle \leftrightarrow \left|F=3,M=+1\right\rangle$ transition at $B_0 = \SI{288}{G}$.
The $Q$-factor was chosen to be large enough to increase the achievable field gradient at the ion position, but low enough to obtain a large microwave bandwidth, which enables state preparation in the hyperfine manifold where the most distant frequency is that of the $\left|F=4,M=+4\right\rangle \leftrightarrow \left|F=3,M=+3\right\rangle$ transition at $\approx \SI{2.5}{GHz}$.
The geometry of the microwave electrode was optimised to maximise the microwave field gradient whilst minimising the microwave field strength at the position of the RF null, where the ion is trapped.
The microwave fields were simulated using finite-element analysis software \cite{HFSS, COMSOL} (see Sec.
\ref{sec:FEM}), as at microwave frequencies the proximity of other conductors leads to a complex distribution of currents which cannot be solved analytically.
The microwave electrode is placed between the two arms of the RF electrode, as shown in figure \ref{fig:trap_schematic}, and is surrounded by ground planes to reduce coupling between sections of the microwave electrode.

\begin{figure*}
\includegraphics[width=\textwidth]{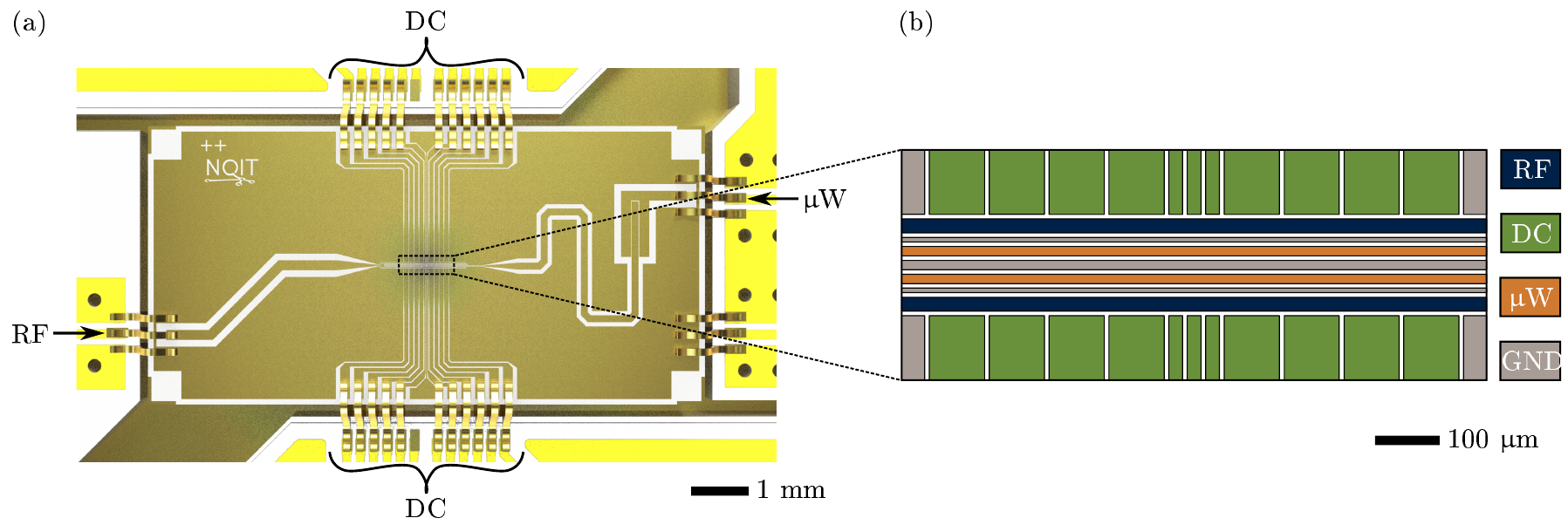}
\caption{%
(a) Rendered image of the surface trap detailing the electrical connections for the radio frequency (RF), DC, and microwave ($\upmu$W) electrodes.
The meander path for the microwave line allows sufficient length to form the quarter-wave resonator.
(b) Enlarged scale view of the central trapping region, showing the structure of the DC electrodes.
}
\label{fig:trap_render}
\end{figure*}

A rendered image of the optimised surface trap design is shown in figure \ref{fig:trap_render}(a), and its central trapping region is detailed in figure \ref{fig:trap_render}(b).
The widths of the electrodes shown in the central trapping region are given in table \ref{tab:electrodes}.
The microwave electrode path to the trap region was chosen to ensure the correct length to form the $\lambda/4$ resonator.
The trapping region has 22 DC electrodes in total, 11 per side, where the central 3 electrode pairs are narrower to facilitate splitting of ion chains.

Microwave power is delivered through a $\SI{50}{\Omega}$ feedline to provide good impedance matching to the microwave source.
This feedline is capacitively coupled to the microwave electrode to inhibit DC coupling to the electrode.
Inevitably, this coupling capacitor also allows microwave energy to leak out of the resonator, reducing the total quality factor of the resonator.
We utilize this fact to limit the quality factor at cryogenic temperatures.
With our expected residual resistivity ratio (RRR) of 15~\cite{private_communication}, the quality factor would change from $\sim 8$ at room temperature to $\sim 120$ (linewidth $\sim50$ MHz) at cryogenic temperatures if only determined by metal resistivity.
The chosen capacitance to the feedline limits the total quality factor to $Q_\text{tot}<17$.

\begin{figure}
\includegraphics[width=0.5\textwidth]{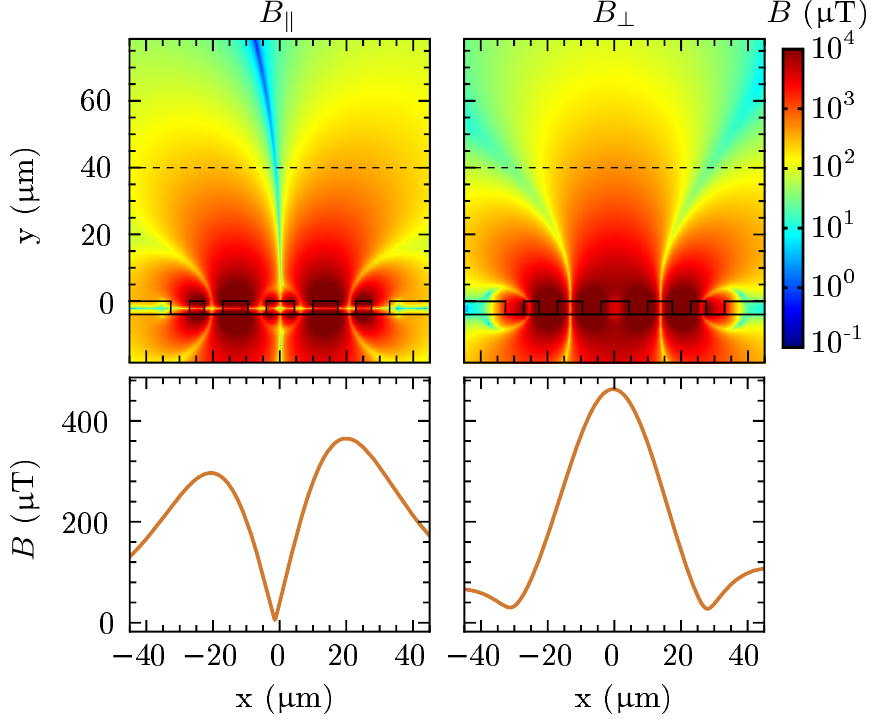}
\caption{Simulated room-temperature magnetic field components parallel $B_{\parallel}$ and perpendicular $B_{\perp}$ to the atomic quantisation axis produced by the microwave electrode at $\SI{3.1}{GHz}$ for an input power of $\SI{1}{W}$.
The upper figures show the magnetic field in a 2D plane above the surface trap ($y=0$), and the lower figures show a slice at the ion trapping height of $y=d=\SI{40}{\micro m}$, where $(x=0,y=40){\SI{}{\micro m}}$ is the position of the RF null.}
\label{fig:bfield}
\end{figure}

The magnetic field components produced by the optimised microwave electrode design, both at room and cryogenic ($T=\SI{20}{K}$) temperatures, are detailed in table \ref{tab:opt_electrode} and figure \ref{fig:bfield}.
The figure shows the simulated magnetic field parallel $B_{\parallel}$ and perpendicular $B_{\perp}$ to the quantisation axis produced by the microwave electrode at $\SI{3.1}{GHz}$ for an input power of $\SI{1}{W}$.
The upper plot shows a 2D plane above the trapping region, and the lower plots show a transect at the ion trapping height of $d=\SI{40}{\micro m}$.
The simulations show that we are able to create a minimum in $B_{\parallel}$, as expected from the simplified model shown in figure \ref{fig:trap_schematic}.
Note that the minimum of the magnetic field is displaced slightly from the RF null ($x=0$) due to the asymmetry in the current amplitude on either side of the ion, resulting in a small residual magnetic field amplitude which couples to the clock qubit (see Sec.
\ref{sec:MW_minimum_simplistic}).
This does however enable the application of a carrier tone which can be used to implement the two-qubit dynamically-decoupled M{\o}lmer-S{\o}rensen (DDMS) gate \cite{Molmer1999, Sorensen1999, Harty2016} and single-qubit rotations.

The quantities calculated at cryogenic temperature assume knowledge of the RRR, which characterizes the change in resistance of the gold trap top-layer (see section \ref{sec:trapfab}) between room and cryogenic temperatures.
The RRR depends strongly on the purity of the gold \cite{Matula1979}, as well as on the deposition methods.
The value quoted above, RRR$=15$, used in simulation and design of the device, is at best an educated guess.
The impact of the uncertain RRR value is mitigated by the relatively large coupling capacitance which is expected to dominate the resonator losses at cryogenic temperatures.

\begin{table}
\begin{ruledtabular}
\begin{tabular}{ccc}
\textbf{Electrode} & \textbf{Design} ($\boldsymbol{\upmu}$\textbf{m}) & \textbf{Measured} ($\boldsymbol{\upmu}$\textbf{m}) \\
\colrule
Dielectric gap & 4.5 & 5.2 \\
Microwave electrode & 8.5 & 7.8\\
RF electrode & 18.8 & 18.4\\
Wide DC electrode & 85.5 & 85.3 \\
Narrow DC electrode & 25.5 & 25.0\\
Inner ground & 9.5 & 8.8 \\
Outer ground & 5.5 & 4.8 \\
\end{tabular}
\end{ruledtabular}
\caption{\label{tab:electrodes} The designed and measured widths of electrodes on the optimised surface trap.}
\end{table} 

\begin{table}
	\begin{ruledtabular}
		\begin{tabular}{cccc}
			\textbf{Temperature} & $\boldsymbol{\partial B_{\parallel}/\partial x}$ \textbf{(T/m)} & $\boldsymbol{B_{\parallel}^0}$ \textbf{(}$\boldsymbol{\upmu}$\textbf{T)} & $\boldsymbol{B_{\perp}^0}$ \textbf{(}$\boldsymbol{\upmu}$\textbf{T)}\\
			\colrule
			$\SI{300}{K}$ & $\num{26}$ & $\num{39}$ & $\num{462}$\\
			$\SI{20}{K}$ & $\num{72}$ & $\num{108}$ & $\num{1282}$\\
		\end{tabular}
	\end{ruledtabular}
	\caption{\label{tab:opt_electrode}Simulated microwave field gradient, and field amplitude parallel and perpendicular to the quantisation axis defined by $\boldsymbol{B_0}$ at the RF null for an input microwave power of $\SI{1}{\watt}$.
For a perfectly symmetric design, we would have $B_{\parallel}^0=0$; the imperfect nulling of $B_{\parallel}^0$ is useful in practice for driving the qubit carrier transition.}
\end{table}

\section{Surface trap fabrication}
\label{sec:trapfab}

The surface trap was constructed from a $\SI{430}{\micro m}$ thick, high-purity ($\SI{99.996}{\percent}$), mono-crystalline sapphire wafer onto which layers of Ti ($\SI{40}{nm}$), Pt ($\SI{60}{nm}$), and Au ($\SI{100}{nm}$) were sequentially deposited using electron-beam evaporation.
A sapphire substrate was chosen for its low loss tangent and excellent thermal conductivity at cryogenic temperatures.
The Ti creates an adhesion layer and the Au provides a chemically passive surface finish.
At the elevated temperatures used to bond the surface trap to the cryogenically-cooled pillbox (see section \ref{sec:cryo}), the Ti would start to diffuse into the Au layer \cite{Martinez2010}; the Pt layer acts as a diffusion barrier between the Ti and Au surfaces \cite{Li2017}, ensuring a pure Au finish.
To create the surface trap electrodes, a $\SI{6}{\micro m}$ photo-resist was applied to the top side, and then selectively exposed to UV light through a mask.
Photo-resist which was exposed to the UV light was removed, after which a layer of Au was electroplated onto the surface trap (see \cite{Wolf2019} for details of the electroplating process).
The Au layer was chosen to be $\SI{4}{\micro m}$ thick as this is several times thicker than the room temperature skin depth of $\SI{1.35}{\micro m}$ at $\SI{3.1}{GHz}$.
The remaining photo-resist was then also removed, and the metallic layers beneath the photo-resist were erased using argon-ion milling, revealing the finished surface trap.
Approximately 64 trap chips were made on the same 3'' sapphire wafer, with some design variations (for example, with a range of microwave resonator lengths so that the chip whose resonator most closely matched the qubit frequency could be selected); the wafer was diced into individual chips.

\section{Filter boards}
\label{sec:filterboards}

The surface trap was electrically connected via gold wire-bonds to several printed-circuit boards (PCBs).
These PCBs were manufactured from alumina substrates ($\SI{96}{\percent}$ purity), with $\SI{60}{\micro m}$ copper traces, coated with a $\sim\SI{100}{nm}$ layer of electroless palladium autocatalytic gold (EPAG).
Alumina has good thermal conductivity and low loss tangent, and EPAG provides a nickel-free, gold surface finish which facilitates Au-Au wire-bonding.
The undersides of the PCBs were also EPAG coated to provide a ground plane connection to the pillbox (figure~\ref{fig:chamber_render}).
All electrical components on the PCBs were soldered using SAC305 solder as it is non-magnetic and ultra-high-vacuum (UHV) compatible.
UHV compatible flux \footnote{P/N: 110797, Accu-Glass Products, Inc.} was also used to improve the quality of solder joints.

There are five PCBs in total: two DC PCBs, a microwave PCB, an RF PCB, and a temperature sensor PCB.
Each DC PCB contains eleven separate traces with an integrated single-pole RC filter ($f_c = \SI{160}{kHz}$) on each trace.
The cut-off frequency $f_c$ was chosen to be significantly lower than the typical secular frequencies of trapped ions (several MHz), but high enough to facilitate rapid ion transport operations.
The RF and microwave PCBs are both singular, grounded coplanar waveguides which were designed to be $\SI{50}{\Omega}$ impedance-matched at their respective frequencies.
Both of these PCBs have multiple vias between the top and bottom ground planes, close to the waveguides to reduce the propagation of parasitic modes.
The RF PCB also contains a $\sim\SI{1}{nH}$ inductive choke in series which was intended to reduce unwanted microwave coupling.
Finally, the temperature sensor PCB contains a Kelvin-connected resistance temperature detector (RTD), and is mounted on the side of the pillbox (figure~\ref{fig:chamber_render}).

\section{Cryogenic Design} 
\label{sec:cryo}

There have been multiple studies \cite{Labaziewicz2008, Deslauriers2006, Daniilidis2011, Chiaverini2014, Noel2019} showing the benefits of operating ion traps at cryogenic temperatures.
The data displayed in \cite{Chiaverini2014} suggests that cooling from room temperature to $T\sim\SI{20}{K}$ reduces the motional heating rate $\dot{\bar{n}}$ by nearly two orders of magnitude (with only a factor $\sim 2$ further reduction available by cooling further to $T\sim\SI{4}{K}$).
Furthermore, cryogenic pumping leads to extremely low background-gas pressures, which can reduce the trapped ion loss rate to negligible levels \cite{Diederich1998}.
In addition to these benefits, the cryogenic temperature reduces the resistive losses in the microwave electrode and thus increases the quality factor of the resonator, leading to a larger field gradient for a given input power.
The thermal conductivity of sapphire is maximal near \SI{20}{K}, allowing efficient transfer of heat from the microwave electrode to the pillbox.

\begin{figure}
\includegraphics[width=\columnwidth]{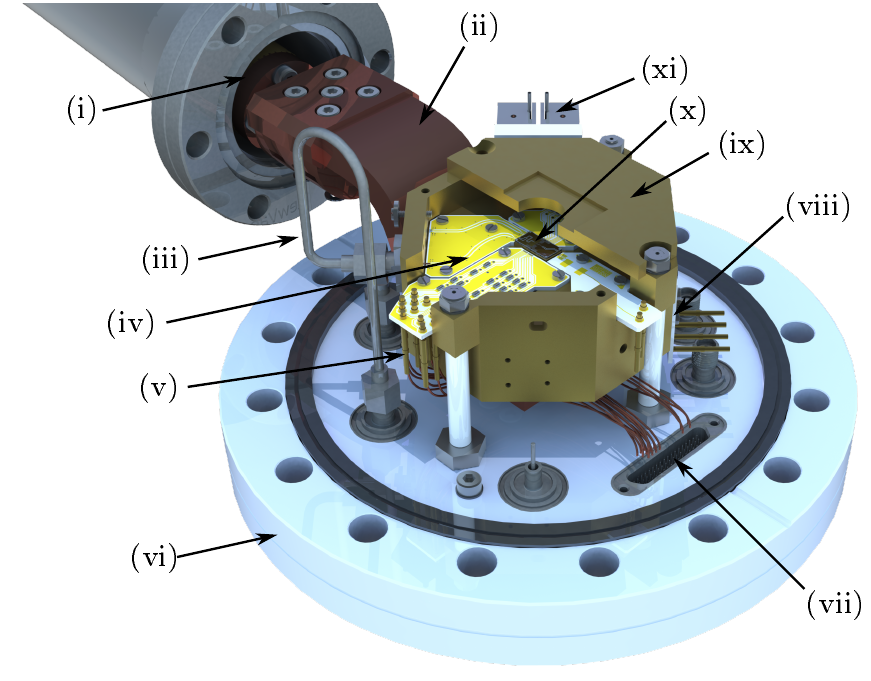}
\caption{Rendered image of the vacuum chamber.
Labelling is as follows: (i) LHe flow cryostat head; (ii) thermal strap; (iii) microwave semi-rigid coaxial cable; (iv) microwave PCB; (v) DC electrode connections; (vi) CF100 base flange; (vii) 51 pin micro-D feedthrough; (viii) temperature sensor; (xi) pillbox lid (cut away for illustration); (x) surface trap chip; (xi) atomic ovens \cite{Ballance2018}.}
\label{fig:chamber_render}
\end{figure}

There are two main challenges when designing a cryogenically-cooled apparatus.
Firstly, the surface trap must be sufficiently thermally isolated from its surrounding room temperature vacuum chamber that it can be effectively cooled with reasonable power, whilst still maintaining good optical access and electrical connections.
Secondly, one must ensure that the cryostat is mechanically isolated from the surface trap to minimise motion of the trap with respect to the optical beam-delivery components.

There are many possible cryostat designs which can be used.
We opted for a Janis ST-400 continuous-flow cryostat which does not require a complex vibration isolation stage.
The apparatus design was optimised to reduce the passive heat load, whilst ensuring an excellent thermal contact between the surface trap and the cryostat.
A thermal budget of the system is shown in table \ref{tab:thermal_budget} \cite{Wolf2019}.
Figure \ref{fig:chamber_render} shows a rendered image of the base flange of the vacuum chamber and the cryostat.
At the centre of the vacuum chamber sits an oxygen-free high-conductivity (OFHC) copper pillbox mounted on three thin-walled ($\SI{0.24}{mm}$) stainless steel legs.
The surface trap and relevant PCBs are mounted inside the pillbox.
The pillbox provides a strong mechanical structure and large thermal mass for the surface trap.
The large thermal mass is particularly important when driving large currents for microwave-driven gates.
The pillbox also creates a low pressure (cryo-pumped) environment around the surface trap when operating at cryogenic temperatures, providing an enhancement in ion lifetime.

\begin{table}
\begin{ruledtabular}
\begin{tabular}{cccc}
\textbf{Source} & \textbf{Material} & \textbf{Power (W)} \\
\colrule
Support legs & 316 stainless steel & 1.43 \\
DC wires & BeCu & 0.18 \\
RF and GND wires & BeCu & 0.08 \\
RTD wires & BeCu & 0.04 \\
$\upmu$W coax & St.St./OFC/PTFE/Ag & 0.13 \\
Radiative & --- & 0.15 \\
\colrule & \textbf{Total} & \textbf{2.01} \\
\end{tabular}
\end{ruledtabular}
\caption{Expected thermal power ingress to the pillbox via vacuum chamber components and radiative transfer.}
\label{tab:thermal_budget}
\end{table}

The pillbox was electroplated with a $\sim\SI{5}{\micro m}$ layer of Au to reduce the absorption of black-body radiation, as well as to improve the bonding of the surface trap to the pillbox.
The surface trap was bonded to the centre of the pillbox using eutectic bonding.
A eutectic bonding process was chosen over other epoxy methods as the thermal conductivity of available epoxies can become a bottleneck in the thermalisation between the surface trap and pillbox.
We soldered the surface trap to the pillbox using a $\SI{25}{\micro m}$ thick eutectic solder sheet with a composition of $\SI{80}{\percent}$ Au and $\SI{20}{\percent}$ Sn.
This bonding process was performed {\em in vacuo} to mitigate oxides forming on the surface trap, at a temperature of $\sim\SI{300}{\celsius}$ \cite{Matijasevic1993}.
A constant pressure of $\SI{0.5}{MPa}$ was applied to the surface trap during the bonding process to ensure a mechanically reliable bond.
The PCBs were screw-mounted to the pillbox and electrical connections were made to the base flange.
The DC and RF connections were made using push-fit crimp pins and BeCu wire, the latter used to minimise thermal conductivity.
The microwave connection was made using a UHV-compatible semi-rigid coaxial cable.

Optical access for laser beams is provided by $\SI{1.25}{mm}$ holes cut into the side faces of the pillbox.
The diameter of the holes was minimised to reduce the vacuum conductivity, thereby enhancing differential pumping at cryogenic temperatures.
These holes may also be used to aid laser beam alignment.
A larger diameter hole in the pillbox lid allows collection of ion fluorescence with a numerical aperture of up to NA$=0.6$.
The atomic source is a short response-time resistively-heated oven \cite{Ballance2018}.
This is mechanically isolated from the pillbox to ensure minimal thermal coupling.
The temperature of the pillbox is measured using a Kelvin-connected RTD \footnote{DT-670-SD-1.4L, Lakeshore Electronics} mounted to the side of the pillbox.
An RTD was chosen over a temperature diode as rectification of RF or microwave signals can lead to systematic shifts in the temperature measurement.
  
On cooling to cryogenic temperatures, the cryostat cold head contracts by $\sim\SI{0.9}{mm}$.
To ensure that the trap is mechanically decoupled from this contraction, the pillbox is connected to the cryostat head using a $\SI{60}{mm}$ long, S-shaped thermal strap \footnote{Thermotive LLC}.
The thermal strap is constructed from 205 sheets of $\SI{25}{\micro m}$-thick copper foil, providing a flexible thermal connection with a measured room temperature conductivity of $\SI{0.80 \pm 0.05}{W/K}$.
The surface trap moves a horizontal distance $<\SI{10}{\micro m}$ when cooled from room temperature to $T=\SI{20}{K}$, relative to the laser beam positions, demonstrating the effectiveness of this mechanical isolation.

\section{Experimental Characterization}
\label{sec:results}

\subsection{Electrical measurements of trap chip}

\begin{figure}
\includegraphics[width=\columnwidth]{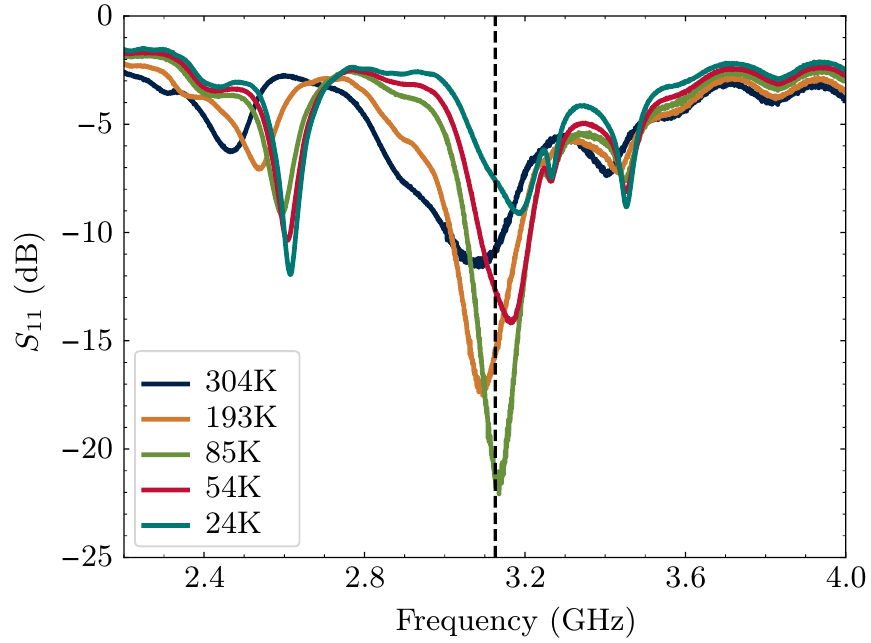}
\caption{Measured $S_{11}$ parameter of the microwave electrode as a function of trap temperature.
The vertical dashed-dotted line shows the clock-qubit frequency at $\left|\mathbf{B_0}\right| = \SI{288}{G}$.}
\label{fig:S_param}
\end{figure}

A vector network analyser (VNA) was used to measure the power reflection ($S_{11}$ parameter) of the microwave electrode as a function of trap temperature.
Results are shown in figure \ref{fig:S_param}.
We observe a broad resonance, which is centred on the qubit frequency at cryogenic temperatures.
We fabricated traps with a variety of resonator lengths, and chose the length that minimized the $\left|S_{11}\right|$ parameter at the qubit frequency at cryogenic temperatures.
The modest quality factor $Q$ of the resonance at room temperature allows the use of the trap without cryogenics, even though the resonant frequency is detuned from the qubit frequency.
Other resonances are also observed, which we attribute to coupling between the microwave and RF electrodes (see Sec.
\ref{sec:MW_minimum}).

There are several characteristic features which change as a function of trap temperature.
The first is the depth of the observed resonance, which is characterised by the quality factor $Q$.
In this microwave resonator, the internal quality factor $Q_\text{int}$ is mostly fixed by resistive losses in the resonant structure, and the external quality factor $Q_\text{ext}$ is determined by the capacitive coupling of the microwave resonator to the feedline \cite{Teufel2011}.
Together these give the overall quality factor $Q$ via $Q^{-1}=Q_\text{ext}^{-1} + Q_\text{int}^{-1}$.
As the trap is cooled, the resistance of the gold decreases, which reduces resistive loss of the microwave resonator and hence increases $Q_\text{int}$.
$Q_\text{ext}$ is approximately constant, as the coupling between the feedline and the microwave resonator is defined by their geometry \footnote{One would expect a small change in coupling, on the order of a few percent, due to a change in dielectric constant of the sapphire substrate; however this is negligible compared to the change in resistive losses of the resonator.}.
At room temperature, the microwave resonator is under-coupled to the feedline ($Q_\text{ext}>Q_\text{int}$).
In this regime, the microwave current which is sent to the microwave electrode will mostly be reflected off the resonator, leading to a loss of power which could potentially have entered the resonator.
As the trap is cooled, the microwave resonator becomes critically-coupled to the feedline ($Q_\text{int}=Q_\text{ext}$), and we observe the smallest $\left|S_{11}\right|$ and maximum power transfer to the resonator.
As the trap is cooled further, the microwave electrode becomes over-coupled ($Q_\text{ext}<Q_\text{int}$), and we observe an increase in $\left|S_{11}\right|$.
Intuitively, when $Q_\text{ext}<Q_\text{int}$, too much energy will be leaking out of the microwave resonator into the feedline, damping the resonator and reducing the amplitude within it.
It is challenging to design a critically-coupled resonator at a given temperature due to the unknown resistance of the gold prior to fabrication.
To optimize the microwave current per unit input power for operation around $\SI{20}{K}$, we would have to increase the external quality factor, by reducing the capacitance between the resonator and the feedline.

Another noticeable change is the shift in frequency of the microwave resonance, to which there are two main contributions.
The dielectric constant of the sapphire changes by approximately $\SI{1.5}{\%}$ between room temperature and liquid helium temperatures \cite{Molla1993}.
This leads to a small change in the in-coupling capacitance as well as to an effective length reduction of the microwave resonator.
The reduction in the gold resistance also contributes to a frequency shift of the resonance (see Sec.
\ref{sec:mw_measurements}).

\subsection{Microwave field measurements}

The microwave-frequency magnetic field produced above the trap chip by the microwave resonator was measured using a single $^{43}$Ca$^+$ ion.
Ions were loaded into the trap using isotope-selective photo-ionisation \cite{Lucas2004} from an isotopically enriched source ($\SI{30}{\%}$ $^{43}$Ca$^+$, $\SI{70}{\%}$ $^{40}$Ca$^+$).
The trapping RF was supplied at a frequency of $\SI{61.3}{MHz}$ and amplified to a voltage of $\approx\SI{80}{V}$ using a series $L-C$ circuit, producing a radial secular frequency of $\approx\SI{5.5}{MHz}$, and radial trap depth of $\sim\SI{30}{meV}$.
The axial secular frequency of $\SI{1.1}{MHz}$ was controlled via application of voltages to the DC electrodes.
The DC electrodes were also used to break the degeneracy of the radial modes by $\sim \SI{0.3}{\mega \hertz}$ and to tilt the radial mode angles by $\SI{15}{\degree}$ to the trap plane, enabling efficient laser cooling with a single laser beam.
Real-time experimental control was provided by modules from the ARTIQ/Sinara open-source ecosystem \cite{artiq}.

Measurements of the microwave field amplitude parallel and perpendicular to the quantisation axis provided by the $\SI{288}{G}$ static magnetic field were made by measuring the Rabi frequency on the $\left|F=3,M=+1\right\rangle \rightarrow \left|F=4,M=+1\right\rangle$ and $\left|F=3,M=+1\right\rangle \rightarrow \left|F=4,M=+2\right\rangle$ transitions respectively.
The ion was initially laser cooled using counter-propagating $\SI{397}{nm}$ and $\SI{866}{nm}$ laser beams using techniques similar to those in \cite{Allcock2016}, after which it was prepared in the $\left|F=4,M=+4\right\rangle$ state via optical pumping with a circularly-polarised $\SI{397}{nm}$ beam propagating parallel to the static magnetic field.
The ion was subsequently prepared in the $\left|F=3,M=+1\right\rangle$ state using a series of microwave $\pi$-pulses (see figure \ref{fig:energy_levels}).
Rabi flopping was then driven on the transition of choice, after which microwave pulses transferred population in the $\left|F=3,M=+1\right\rangle$ state back to the $\left|F=4,M=+4\right\rangle$ state.
Population in the $\left|F=4,M=+4\right\rangle$ state was shelved in the metastable 3D$_{5/2}$ level using $\SI{393}{nm}$ and $\SI{850}{nm}$ laser pulses \cite{Myerson2008}.
The Doppler cooling beams were then applied and the state of the ion inferred by the absence or presence of ion fluorescence.
Finally, the ion was repumped to the ground level using an $\SI{854}{nm}$ pulse.

\begin{figure}
\includegraphics[width=\columnwidth]{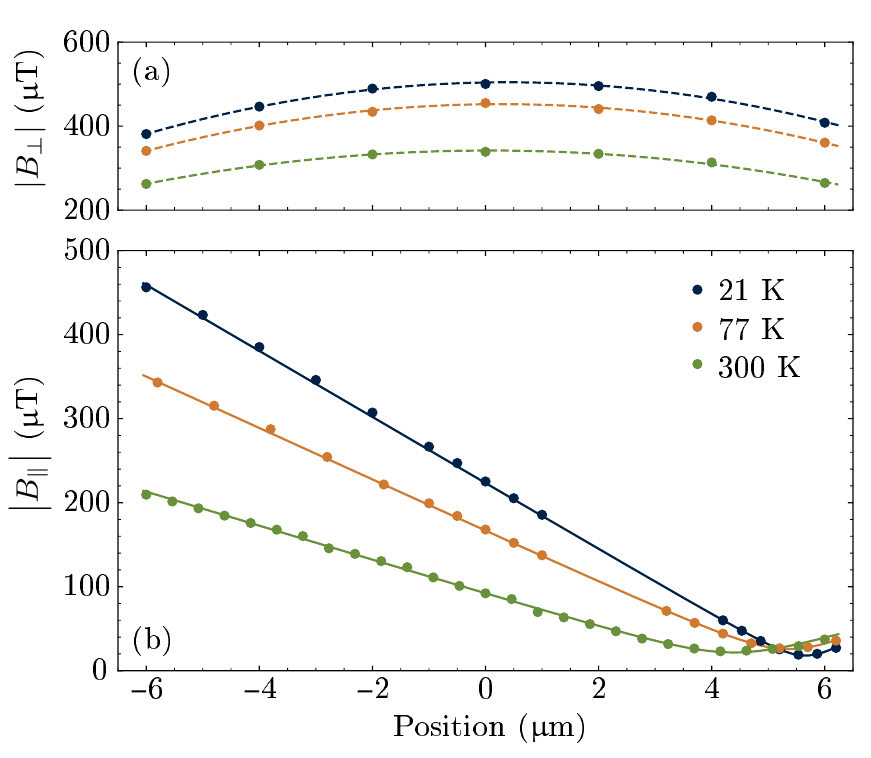}
\caption{%
Microwave field measured with a single ion, for field amplitude (a) perpendicular $B_{\perp}$ and (b) parallel $B_{\parallel}$ to the static field $B_0$, at trap temperatures of $T = \SI{300}{K}$, $\SI{77}{K}$, and $\SI{21}{K}$ (green, orange, blue symbols respectively).
The measured microwave field amplitude is normalised to an input power of $\SI{1}{W}$.
The dashed lines in (a) are guides to the eye, while the solid lines in (b) are fits to the data using equation \ref{eq:grad}.
The trap centre (RF null) is at $x=0$.}
\label{fig:gradient}
\end{figure}

To measure the microwave field amplitude as a function of $\hat{x}$, we varied the ion position by adjusting the trap DC voltages.
The measured spatial distribution is shown in figure \ref{fig:gradient}, at trap temperatures of $T = \SI{300}{K}$, $\SI{77}{K}$, and $\SI{21}{K}$.
As discussed previously, one expects a minimum in the component of the microwave field parallel to the static magnetic field (see figure \ref{fig:trap_schematic}(b)), resulting in a microwave field with a large gradient but small amplitude.
As expected, we observe a minimum in this microwave component; however, this minimum is offset by several microns from the RF null.
Consequently, the parallel microwave amplitude component $B_\parallel$ at the trap centre is larger than designed.
Through further simulations and measurements, we deduced that this is caused by a resonant coupling between the microwave and RF electrodes (see Sec.
\ref{sec:MW_minimum}), due to an impedance mismatch at a boundary formed by the RF choke, its large solder pads, and the $\sim\SI{200}{\ohm}$ RF-carrying wire.
The impact of the choke on the microwave reflection is small relative to that of the RF wiring and the capacitance ($>\SI{1}{\pico \farad}$) of the solder pads.
We estimate a $\approx \SI{70}{\%}$ reflection of the coupled microwave power from these components.
This effect can be completely mitigated in future surface traps by creating a well-defined impedance boundary for microwaves on the RF line, such that a resonance cannot be formed.
Simulations show that a spiral inductor, in line with the RF electrode, and placed at a distance $\lambda/4$ from the RF electrode ends (see figure~\ref{fig:Choke}(b,c)), has the desired effect.

By considering the simplified model in figure \ref{fig:trap_schematic}(b), the complex amplitude of the microwave field parallel to the static magnetic field can be approximated by 
\begin{equation} \label{eq:grad}
B_{\parallel}\left(x\right) = B_{\parallel}^{x_0} + {\rm{e}}^{\rm{i}\varphi}\frac{\partial B_{\parallel}}{\partial x}(x-x_0)
\end{equation}
where $B_{\parallel}^{x_0}$ is the residual magnetic field amplitude at $x=x_0$, the position of the microwave minimum, the RF null is at $x=0$ and $\varphi$ is the phase shift between the microwave amplitude and gradient resulting from the phase difference of the opposing currents in the microwave electrode.
The fitted gradients and field amplitudes at the RF null are displayed in table \ref{tab:gradient_fit}.
We observe an increase in the measured field gradient as the trap temperature is reduced, consistent with the decrease in the resistivity of the gold surface layer of the trap.
From a previous iteration of the surface trap, which did not suffer from resonant coupling to the RF electrode, we were able to extract the resistivity of the gold as a function of trap temperature (see Sec.
\ref{sec:mw_measurements}), and hence $Q$.
The temperature scaling of $Q$ is consistent with that of the measured microwave field gradient.

From the measured field gradient and field amplitude we are able to calculate an effective Lamb-Dicke parameter $\eta$ at the RF null, defined as $\eta = q_0\partial_x B_{\parallel}/B_{\parallel}^0$ where $q_0$ is the r.m.s.\ extent of the ion's ground-state wave packet.
At $\SI{20}{K}$ and a radial secular frequency of $\SI{5.5}{MHz}$, $\eta \approx \num{8e-4}$, which is comparable to $\eta \approx \num{1e-3}$ in our previous work \cite{Harty2016}.
If the microwave minimum were at its designed position, we would obtain $\eta \approx \num{3e-3}$. Inspite not achieving an increased effective Lamb-Dicke parameter, we are able to achieve significantly faster two-qubit gates.
   
\begin{table}
\begin{ruledtabular}
\begin{tabular}{ccc}
\textbf{Temperature} & $\boldsymbol{\partial B_{\parallel}/\partial x}$ \textbf{(T/m)} & $\boldsymbol{B_{\parallel}^0}$ \textbf{(}$\boldsymbol{\upmu}$\textbf{T)}\\
\colrule
$\SI{300}{K}$ & $\num{20.4(2)}$ & $\num{92.2(4)}$\\
$\SI{77}{K}$ & $\num{30.7(2)}$ & $\num{166.8(5)}$\\
$\SI{21}{K}$ & $\num{39.4(2)}$ & $\num{223.4(9)}$\\
\end{tabular}
\end{ruledtabular}
\caption{\label{tab:gradient_fit}%
Measured microwave field gradient along the $\hat{x}$ direction, and field amplitude at the RF null, at different trap temperatures with a microwave input power of $\SI{1}{\watt}$, for the $B_\parallel$ field component that couples to the qubit transition.}
\end{table}
 
\subsection{Measurements of ion trapping lifetimes and motional mode heating rates}
 
One of the advantages of operating at cryogenic temperatures is an increase in ion trapping lifetime due to a reduction of background gas pressure in the UHV chamber.
During room temperature operation, the UHV chamber vacuum pressure was measured with an ion gauge to be $\approx \SI{1e-11}{mbar}$.
The approximate single- and two-ion trapping lifetimes with continuous Doppler cooling are displayed in table \ref{tab:lifetime}.
We observe a rapid increase in ion lifetime as the temperature of the pillbox is reduced.
At cryogenic temperatures, the saturation pressure of typical gases in a UHV chamber fall below $\SI{1e-12}{mbar}$ \cite{Honig1960}, resulting in a reduction in local background gases near the trapped ions.
The dominant source of residual background gas at $\SI{21}{K}$ is expected to be hydrogen.
This could also be reduced by cooling below $\SI{16}{K}$, but the cost of cryogen outweighs the benefits given the observed two-ion lifetime.
Also, the pillbox has multiple apertures which allow entry of other background gases from the room temperature region of the chamber.

\begin{table}
\begin{ruledtabular}
\begin{tabular}{ccc}
\textbf{Temperature} & $\boldsymbol{\tau_1}$ \textbf{(minutes)}& $\boldsymbol{\tau_2}$ \textbf{(minutes)}\\
\colrule
$\SI{300}{K}$ & $\approx 30$ & $\approx 3$ \\
$\SI{77}{K}$ & $>120$ & $\approx 20$\\
$\SI{21}{K}$ & $>600$ & $>120$\\
\end{tabular}
\end{ruledtabular}
\caption{\label{tab:lifetime}Approximate single-ion ($\tau_1$) and two-ion ($\tau_2$) trapping lifetimes with continuous Doppler cooling, as a function of surface trap temperature.}
\end{table}

Another advantage of cryogenic operation is the reduction in the heating rate of the motional modes of the ions.
From previous experiments which have characterised ion heating rates as a function of temperature \cite{Labaziewicz2008, Deslauriers2006, Daniilidis2011, Chiaverini2014, Noel2019}, one would expect a power-law relationship of the type $\dot{\bar{n}} \propto T^{\beta}$, where $\beta$ typically varies between 1.5 and 4.

We measured the ion heating rates on the single- and two-ion radial motional modes as a function of surface trap temperature using pulsed sideband cooling \cite{Monroe1995b} and sideband thermometry.
These were implemented by driving the $\left|F=4,M=+4\right\rangle\rightarrow\left|F=4,M=+3\right\rangle$ optical Raman transition and its sidebands.
The two Raman beams were derived from the same laser, detuned $\SI{30}{GHz}$ from the $\SI{397}{nm}$ $4\text{S}_{1/2}\rightarrow 4\text{P}_{1/2}$ transition, and had orthogonal linear polarisations.
The beams propagated along the $\left(\hat{z} + \hat{x}\right)/\sqrt{2}$ and $\left(\hat{z} - \hat{x}\right)/\sqrt{2}$ directions so that there was no Raman coupling to the axial motional modes.
The ions were initially Doppler cooled, after which dark-resonance cooling was applied for a further $\SI{10}{ms}$ \cite{Allcock2016}.
The ions were then prepared in the $\left|F=4,M=+4\right\rangle$ state via optical pumping with the circularly-polarised $\SI{397}{nm}$ beam.
Ground-state cooling was then performed using a pulsed-sideband cooling technique which consists of a repeated sequence of driving the first red motional sideband with the Raman beams, after which the circularly-polarised $\SI{397}{nm}$ beam was used to provide dissipative repumping into the $\left|F=4,M=+4\right\rangle$ state.
Thermometry was then performed by measuring the ratio of the amplitude of the first red and blue motional Raman sidebands.

The measured heating rates for the single-ion in-plane radial mode $\dot{\bar{n}}_1$ and the two-ion in-plane radial rocking mode $\dot{\bar{n}}_2$ are shown in table \ref{tab:heating_rate}.
The mode of most interest is the two-ion in-plane radial rocking mode, as this is the mode on which two-qubit entangling gates are to be performed.
We observe a reduction in the heating rate as a function of temperature with an exponent $\beta \approx 0.2$.
This is a much smaller effect than one would expect from previous cryogenic experiments, and may indicate that the electric field noise arises from technical sources which are currently unidentified.

\begin{table}
\begin{ruledtabular}
\begin{tabular}{ccc}
\textbf{Temperature} & $\boldsymbol{\dot{\bar{n}}_1}$ \textbf{(quanta/s)} & $\boldsymbol{\dot{\bar{n}}_2}$ \textbf{(quanta/s)}\\
\colrule
$\SI{300}{K}$ & $\num{350(50)}$ & --- \\
$\SI{77}{K}$ & $\num{230(20)}$ & $\num{5.5(7)}$\\
$\SI{21}{K}$ & $\num{200(10)}$ & $\num{3.1(9)}$\\
\end{tabular}
\end{ruledtabular}
\caption{\label{tab:heating_rate}Measured heating rates for the single-ion in-plane radial mode $\dot{\bar{n}}_1$ and for the two-ion in-plane radial rocking mode $\dot{\bar{n}}_2$ as function of surface trap temperature.}
\end{table}

\section{Conclusions}

In this paper we have presented a novel surface-electrode ion trap with an integrated microwave resonator, designed to improve upon the speed and fidelity of two-qubit entangling operations presented in our previous work \cite{Harty2016}.
One of the main aims was to increase the achievable microwave-frequency magnetic field gradient for a given input power, whilst maintaining or improving upon the effective Lamb-Dicke parameter $\eta$ and motional-mode heating rate.
With the use of a passively-nulled resonant microwave electrode design and a $\pi$-polarized qubit transition that couples more efficiently to the microwave field, we have obtained a microwave field gradient which is effectively a factor of eight greater than that of our previous system, for the same input power.
This should enable an order-of-magnitude improvement in the speed of two-qubit entangling operations for reasonable input microwave powers \footnote{Microwave input power of a few Watts is easily generated; the trap chip has been tested for powers up to $\SI{16}{W}$ without damage.}.

We have observed a detrimental resonant coupling between the microwave and RF electrodes, which does not affect the achievable gradient, but causes a shift in the position of the microwave field minimum, which in turn reduces $\eta$; however, the measured $\eta$ is comparable to that of our previous work.
We have shown in simulation that this resonant coupling between the microwave and RF electrodes could be mitigated by placement of an inductive choke on the RF electrode at a distance $\lambda/4$ from the microwave resonator.

The ability to cool the surface trap to cryogenic temperatures has yielded ion trapping lifetimes of several hours.
In addition, we have observed a reduction in motional heating rates at cryogenic temperatures, which has enabled us to maintain a comparable motional heating rate to that in our previous system, despite the significantly reduced ion-to-electrode distance.
Future work will utilise these improvements to demonstrate microwave-driven two-qubit entangling operations which are comparable in both speed and fidelity to those of state-of-the-art laser-driven gates.

\section{Acknowledgements}

We thank members of the NIST Ion Storage group, in particular David Allcock and Raghu Srinivas, for much helpful advice on cryogenic ion trap design and operation.
TPH would like to thank NIST for their hospitality during an extended visit.
CJB and TPH are Directors of Oxford Ionics Ltd.
This work was supported by the U.S.\ Army Research Office (ref.
W911NF-18-1-0340) and the U.K.\ EPSRC Quantum Computing and Simulation Hub.

\bibliographystyle{unsrt}
\bibliography{bibfile} 
\clearpage
\onecolumngrid


\begin{center}
{\large\textbf{Supplementary information}}
\vspace{20pt}
\end{center}
\makeatletter
   \renewcommand\l@section{\@dottedtocline{2}{1.5em}{1em}}
   \renewcommand\l@subsection{\@dottedtocline{2}{3.5em}{1em}}
   \renewcommand\l@subsubsection{\@dottedtocline{2}{5.5em}{1em}}
\makeatother

\renewcommand{\theequation}{S\arabic{equation}}
\renewcommand{\thefigure}{S\arabic{figure}}
\renewcommand{\thetable}{S\arabic{table}}
\renewcommand{\thesection}{S\arabic{section}}
\setcounter{figure}{0}
\setcounter{equation}{0}
\setcounter{section}{0}

\twocolumngrid

\section{Finite-element simulation methods}
\label{sec:FEM}

Microwave finite element simulations of the device were carried out in COMSOL Multiphysics \cite{COMSOL}.
Here we describe our approach to these simulations.

We note that the simulated microwave behaviour was strongly dependent on our approach to grounding the surface trap.
As it is unfeasible to simulate a large volume of the experiment (the PCBs, the pillbox, etc.), to improve the accuracy of our ground model we restricted our analysis to the surface trap chip and used an idealized ground.
For this reason, we are cautious about the accuracy of quantitative predictions of the simulation; however, the qualitative results we extract from our simulations were found to be robust to different simulation choices (e.g.\ simulated area or grounding method).

The simulation was constructed by enclosing a model of the surface trap (see figure \ref{fig:Choke}(a)) in a perfectly conducting metal box, which formed the boundary condition of our simulation volume.
The sides of the surface trap were grounded through a connection to this metal box.
We found that the boundary condition on the DC lines had little influence on the microwave behaviour of the surface trap, and these were therefore left grounded.
The conductive layer was modelled as a $\SI{4}{\upmu m}$ thick layer of gold with the resistivity of bulk gold at room temperature: $\SI{22}{n\Omega.m}$~\cite{Matula1979}.
In the vertical ($y$) direction, the simulation volume extended from the bottom of a $\SI{500}{\micro m}$ thick volume of dielectric, with the relative dielectric constant of sapphire averaged over all axis orientations ($\epsilon_\text{sap}\approx 10\epsilon_0$ following supplier specifications), to the top of a $\SI{4.5}{mm}$ thick volume of vacuum.

The boundary conditions on the microwave and RF lines were a lumped $\SI{50}{\Omega}$ microwave source and a lumped complex load respectively (shown in figure~\ref{fig:Choke}(c)).
The latter was chosen to determine qualitatively the impact of an abrupt boundary condition on the RF line, without attempting to model the choke, its solder pads, the high-impedance RF wiring (shorted at microwave frequencies through a capacitor outside the vacuum chamber), and the complicated ground path back to the chip.
The complex load of the RF line was thus simulated as a length of waveguide with a low characteristic impedance of $\SI{50}{\Omega}$ (present on the surface trap, and continued via wire-bonds onto the RF PCB), which transitions to a very high characteristic impedance waveguide.
We modelled this as a $Z_{50} =\SI{50}{\Omega}$ impedance waveguide of length $l$ with an open termination.
The input impedance as a function of length is given by $-jZ_{50}/\tan(-2\pi l/\lambda)$, were $j$ indicates the imaginary number \cite{Pozar2009}.
The magnetic field simulations shown in figure \ref{fig:bfield} were performed using this model, with $l=0$.

\section{On the position of the microwave minimum}
\label{sec:MW_minimum}
\subsection{Microwave minimum in a simplistic model}
\label{sec:MW_minimum_simplistic}

The simulated microwave-frequency magnetic field produced by the surface trap (see figure \ref{fig:bfield}) has a minimum which is slightly displaced from the symmetry axis of the surface trap.
This offset is due to an asymmetry in the microwave current delivery.
The microwave resonator is folded around the ion, such that one side of the resonator is closer to the short to ground than the other.
The change in amplitude and phase of the oscillating current, as a function of distance to the short, then leads to one side of the resonator generating a magnetic field which is slightly different to the other.

\begin{figure}
\includegraphics[width=\columnwidth]{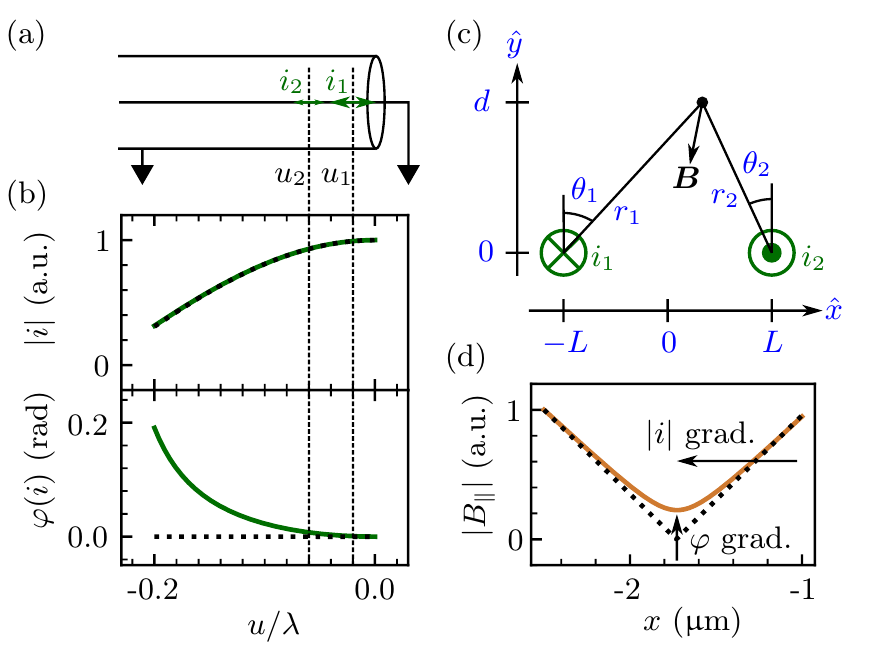}
	\caption{%
(a) Schematic of a short-circuited waveguide which is used to explain the position of the microwave minimum.
(b) Amplitude and phase of the current phasor $i = |i|e^{j\varphi(i)}$ in the waveguide, as a function of position $u$ with respect to the short, for a lossy (solid line) and loss-less (dashed line) conductor.
Vertical dashed lines correspond to the positions along the waveguide which are adjacent to the ion.
(c) Simplified two-wire model used to quantify the impact of the microwave current asymmetry on the position of the minimum.
The forwards and backward moving currents $i_1$ and $i_2$ give rise to the magnetic field \textbf{B}.
Geometric parameters are noted in blue.
(d) $B_{\parallel}$ as a function of position, calculated from the simplified model with lossy (full line) and loss-less (dashed line) conductor.
}
\label{fig:null_shift_schematic}
\end{figure}

Using the simplistic model of a shorted waveguide shown in figure \ref{fig:null_shift_schematic}(a), we can estimate the magnitude of the resulting shift in the magnetic field minimum.
Following \cite{Pozar2009}, the current phasor in a waveguide at a distance $u$ from a short-circuit is given by
\begin{equation}
	i(u) = i(0)\left(e^{-\gamma u}+e^{\gamma u}\right)~,
\end{equation}
where
\begin{equation}
	\gamma = \frac{2\pi}{\lambda}\left(\frac{1}{2Q_\text{tot}}+j\right)~,
\end{equation}
$\lambda$ is the microwave wavelength, and $Q_\text{tot}$ is the total quality factor that a $\lambda/4$ resonator would have if constructed from this waveguide.
The current phasor, as a function of $u$, is plotted in figure \ref{fig:null_shift_schematic}(b).
The wavelength is 
\begin{equation}
	\lambda = \frac{1}{\sqrt{\epsilon\mu_0}}\frac{2\pi}{\omega_r}~,
	\label{eq:wavelength}
\end{equation}
where $\omega_r$ is the microwave angular frequency, and $\epsilon = (\epsilon_0+\epsilon_\text{sap})/2$ is the effective dielectric constant, an average of the vacuum $\epsilon_0$ and sapphire $\epsilon_\text{sap}$ dielectric constants.
For simplicity, we use the average dielectric constant of sapphire over all crystal directions $\epsilon_\text{sap}\approx 10\epsilon_0$ (following values of our wafer supplier).
The wavelength at the clock qubit frequency ($\omega_r = 2\pi\times \SI{3.12}{GHz}$) is $\lambda=\SI{4}{cm}$.
The distances to the grounded end of the resonator from the two waveguide locations closest to the ion are $u_1=-0.019 \lambda$ and $u_2=-0.056 \lambda$.
The microwave currents at these points are $i_1 = i(u_1)$ and $i_2 = i(u_2)$ respectively.

We now use the simplified model shown in figure \ref{fig:trap_schematic}(b), as well as the currents $i_1$, $i_2$, to estimate the microwave minimum position.
This model is shown again in figure \ref{fig:null_shift_schematic}(c), with additional annotations.
The centre conductor of the resonator is represented by two wires separated by $2L=\SI{30}{\upmu m}$ with counter-propagating currents $i_1$ and $i_2$ on each side of an ion, which is located at a height $d=\SI{40}{\upmu m}$ above the trap surface.
Following the Biot-Savart law, these currents will generate a horizontal magnetic field given by $B_{\parallel} \propto i_1 \cos(\theta_1) /r_1 - i_2 \cos(\theta_2) /r_2 = i_1 \cos^2(\theta_1) /d - i_2 \cos^2(\theta_2) /d$.
Here $r_{1,2}$ is the distance from the wire to the ion, and $\theta_{1,2}$ its angle with respect to the vertical direction (see figure \ref{fig:null_shift_schematic}(c)).
The angles are $\theta_1 = \text{arctan}[(L+x)/d]$, $\theta_2 = \text{arctan}[(L-x)/d]$, where $x$ is the horizontal position of the ion with respect to the symmetry axis and RF null.

We use these results to plot $B_{\parallel}$ as a function of position in figure \ref{fig:null_shift_schematic}(d).
For a loss-less conductor ($Q_\text{tot}\rightarrow \infty$), the magnetic field exhibits a null, displaced from the position of the RF null.
By introducing losses ($Q_\text{tot}=10$), we find that the null becomes a broadened minimum at the same location (figure \ref{fig:null_shift_schematic}(d)).
As shown in figure \ref{fig:null_shift_schematic}(b), resistance leads to a phase gradient in the waveguide, in addition to the amplitude gradient.
The phase-gradient is thus responsible for the non-zero minimum, whereas the amplitude gradient leads to its shift in position.
Note that in this simplistic model, the shift exceeds somewhat that simulated in COMSOL; a difference is however expected given that many aspects have been neglected in the simplistic model, such as return currents flowing through the nearby ground plane.

\subsection{Influence of the RF line boundary condition}

\begin{figure}
\includegraphics[width=\columnwidth]{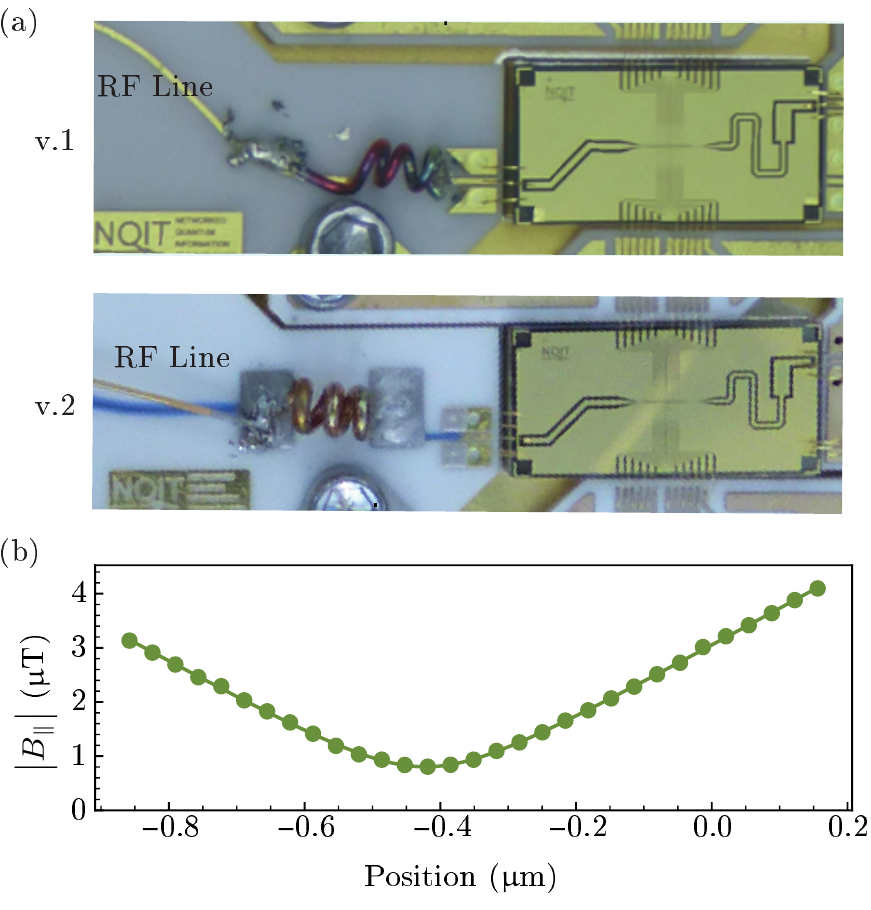}
\caption{%
(a) A photograph of the first (v.1) and second (v.2) iteration of the surface trap, illustrating the change in the RF PCB layout between versions.
(b) Measured $B_{\parallel}$ as a function of position in the v.1 iteration, at a trap temperature of $T=\SI{300}{K}$.
The trap centre (RF null) is at $x=0$.
Compare with the v.2 measurements in figure~\ref{fig:gradient}(b), noting the change of scale on the $x$-axis.
}
\label{fig:Null_shift}
\end{figure}

\begin{figure*}
\includegraphics[]{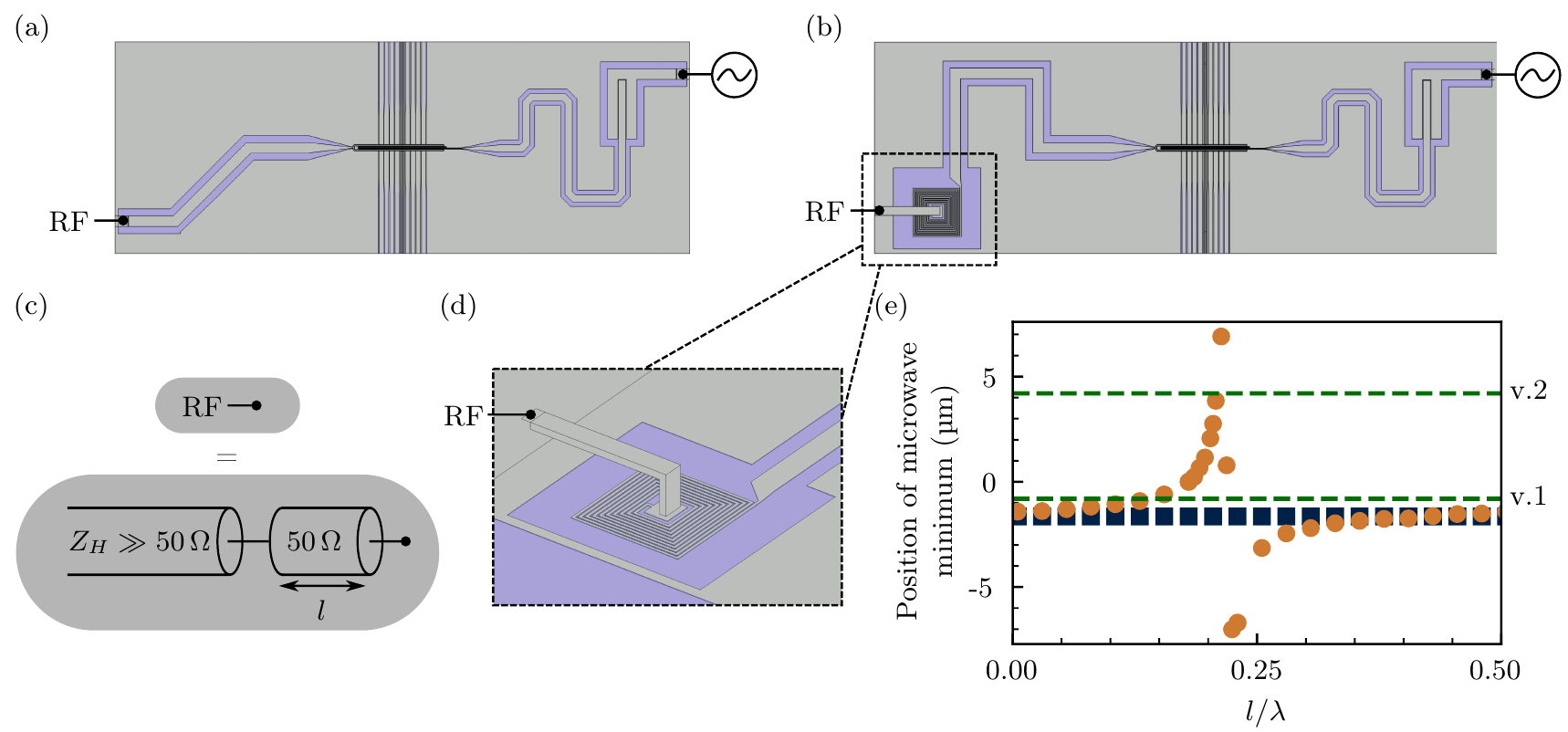}
\caption{
(a) and (b) show top views of the models used in COMSOL simulations.
Grey surfaces are conductive, whilst black/blue surfaces show the dielectric substrate.
Model (a) reproduces our experiment, whilst model (b) includes a planar spiral inductor a quarter microwave wavelength from the RF electrodes.
(c) Details of the RF boundary condition, simulating an impedance mismatch occurring further up the RF line, at a distance $l$ from the edge of the model.
(d) Zoom-in on the planar inductor (isometric view).
The conductive structure connecting the centre of the inductor to the RF boundary condition represents a wire-bond.
(e) Position $x_0$ of the microwave field minimum as a function of $l$ for model (a) (circles) and model (b) (squares).
The dashed horizontal lines show the positions of the measured microwave field minimum for the first (v.1) and second (v.2) iterations of the experimental setup (v.2 is discussed in the main text).} 
\label{fig:Choke}
\end{figure*}

This apparatus has undergone two iterations with different surface traps and PCBs.
In the second iteration, the surface trap was identical except for a slightly different choice of resonator length, but the PCB carrying the RF signal to the trap underwent several significant changes (see figure~\ref{fig:Null_shift}).
First, large solder pads were added to facilitate the soldering of the inductive choke coil, meaning that the coil was placed slightly further from the trap chip.
The pad area leads to a total lumped capacitance of $C_\text{pad}\sim \SI{1.5}{\pico \farad}$, and hence a low $\sim \SI{30}{\ohm}$ path to ground at microwave frequencies.
Secondly, a crack appeared in the RF line on the PCB, which was repaired by soldering a wire a few millimetres above the surface of the PCB.
Since the capacitance of this wire to ground is partially through air rather than solely through the alumina of the PCB, it has an impedance of $Z_\text{RF}\sim \SI{200}{\ohm}$.
Together, these two elements lead to an impedance mismatch with respect to the $\sim \SI{50}{\ohm}$ characteristic impedance of the RF-carrying waveguide on the trap chip.
The mismatch reflects a portion $|\Gamma|^2$ of any incident microwave power, given by $|\Gamma|^2=|(Z_{50}-Z_L)/(Z_{50}+Z_L)|^2$ where $Z_{50}=\SI{50}{\ohm}$ and $1/Z_L=1/Z_\text{RF}+j\omega C_\text{pad}$, amounting to $|\Gamma|^2\sim0.7$.
Whilst such a mismatch also existed in the first iteration of the RF PCB --- where the $\sim \SI{50}{\ohm}$ waveguide on the RF PCB was soldered to a bare wire carrying the RF signal from the vacuum feedthrough --- the length of $\SI{50}{\ohm}$ waveguide capacitively coupled to the microwave resonator has changed, and hence the resonance frequency of any cavity created by the impedance mismatch will be different.

The offset between the RF null and the minimum of the microwave field (shown in figure \ref{fig:gradient}) shifted by $\approx\SI{5}{\upmu m}$ with respect to the first iteration of the experiment (see figure \ref{fig:Null_shift}(b); the field was measured in the same way as for figure \ref{fig:gradient}).
Given the correlation between the change in RF PCB and shift of the microwave minimum, we investigated the impact of the RF line boundary condition in finite-element simulations.

As outlined in section \ref{sec:FEM}, we simulated the RF line through a lumped-element boundary condition at the end of the portion of RF line represented in the COMSOL model, see figure \ref{fig:Choke}(a).
This lumped element was characterized by an impedance $-jZ_{50}/\tan(-2\pi l/\lambda)$; a length of $Z_{50}=\SI{50}{\Omega}$ impedance waveguide terminated after a distance $l$ by an open termination.
We simulated the magnetic field distribution whilst varying $l$ in the RF boundary condition.
The orange circles in figure \ref{fig:Choke}(e) show that at a certain distance $l$ the position of the magnetic field minimum drastically changes, reaching values similar to those measured in the second iteration of the experiment.
At shorter distances, as in the first iteration of the experiment, we find the minimum at a position similar to that initially measured.

At $l \approx 0.2\lambda\approx\SI{8}{mm}$, where the position of the minimum seems to diverge, we found (in simulation) that the current traversing the RF electrodes sharply increases to match that in the microwave electrode.
This indicates that the feature corresponds to a resonant coupling between the RF and microwave electrodes.
In order to suppress this resonance, we propose placing a series inductor (an open termination) at a distance $\lambda/4$ from the end of the RF electrodes, suppressing current oscillations at the position where they would be largest in a resonance.

\subsection{Inductive choke to suppress RF line resonances}

Here, we demonstrate (in simulation) how a planar inductor incorporated onto the surface trap can eliminate microwave resonances in the RF line, which have been shown to cause a shift in the magnetic field minimum.
We constructed a second model, shown in figure \ref{fig:Choke}(b), which included a planar spiral inductor at a distance $\lambda/4$ from the RF electrodes, at a current anti-node.
The inductor consists of 10 turns of $\SI{20}{\upmu m}$ wide conductive trace, which can readily be patterned using optical lithography.
Connection from the centre of the spiral inductor to the RF line on the RF PCB can easily be made via wire-bonding.
It has an inductance of $\SI{58}{nH}$, which corresponds to a $\SI{1}{k \Omega}$ complex impedance at $\SI{3}{GHz}$, and has negligible impedance with respect to the total inductance and resistance of the RF line.
Finally, the inductor has self-resonances at $\SI{1.5}{GHz}$ and $\SI{5}{GHz}$ which would not interfere with its role in this setup.
The blue squares in figure \ref{fig:Choke}(e) show that the on-chip inductor eliminates the influence of the RF line on the position of the microwave minimum.

\section{Supplementary microwave measurements and analysis}\label{sec:mw_measurements}

In this section, we discuss the $S_{11}$ measurements taken in the first iteration (v.1) of the system.
These measurements (figure~\ref{fig:s11_fits}(a)) are devoid of a second resonant peak, which appeared in the second iteration of the experiment (figure~\ref{fig:S_param}), and which we believe is due to the resonant coupling to the RF electrode discussed above.
For this reason, we are able to fit more accurately the $S_{11}$ data and extract the resonator's quality factor, leading to a characterization of the metal conductance as the device is cooled.
We also discuss the origin of an observed shift in resonance frequency with temperature.

\subsection{$S_{11}$ fitting model}

\begin{figure}
\includegraphics[width=\columnwidth]{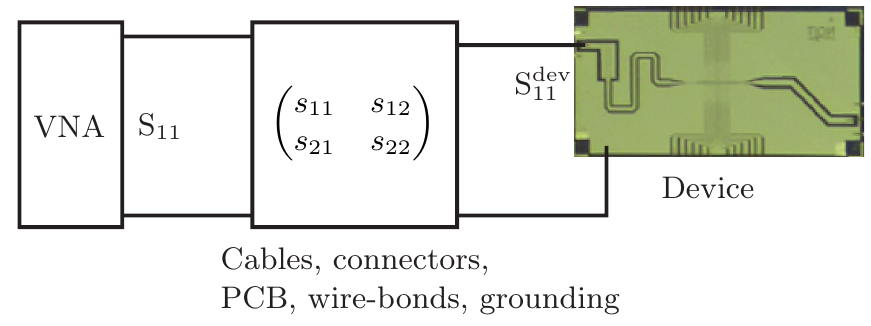}
\caption{Schematic of the microwave setup, including the scattering network $s_{ij}$, situated in the vacuum chamber and separating the VNA from the device under test.}
\label{fig:S11_network}
\end{figure}

\begin{figure}
\includegraphics[width=\columnwidth]{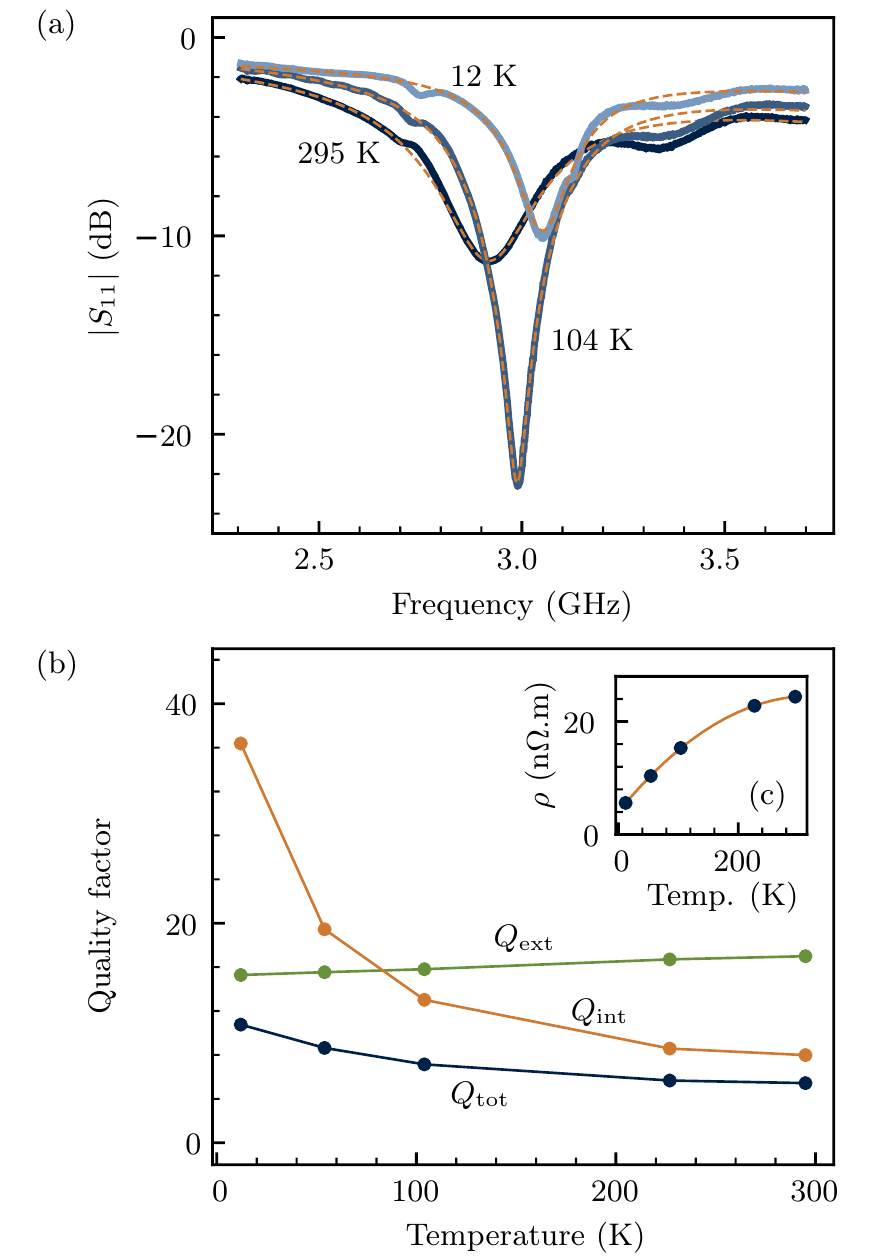}
\caption{
(a) Measured (solid line) and fitted (dashed line) $S_{11}$ at different surface trap temperatures, for the v.1 iteration of the system (compare figure~\ref{fig:S_param} in the main text).
(b) Quality factors extracted from the fits.
(c) Extracted resistivity of the gold layer of the surface trap as a function of temperature.
}
\label{fig:s11_fits}
\end{figure}

In order to accurately fit the $S_{11}$ measurements, we used a model which takes into account the imperfections in the cabling and ground (cables, connectors, PCB, wire-bonds) connecting the VNA to our device.
These were modelled by a two-port network with scattering parameters $s_{ij}$ illustrated in figure \ref{fig:S11_network}.
$s_{11}$ and $s_{22}$ correspond to the reflections at the VNA ports (reflected back to the VNA) and at the device (reflected back to the device).
$s_{21}$ and $s_{12}$ correspond to the attenuation between the VNA and the device.
Following \cite{Pozar2009}, the reflection parameter of our device $S_{11}^\text{dev}$ (coupler and resonator) is given by
\begin{equation}
	S_{11}^\text{dev} = 1-\frac{2Q_\text{tot}}{Q_\text{ext}}\frac{1}{1-2jQ_\text{tot}(\frac{\omega}{\omega_r}-1)}~.
	\label{eq:S11_dev}
\end{equation}
Here $Q_\text{tot} = 1/(Q_\text{int}^{-1}+Q_\text{ext}^{-1})$ is the total quality factor which provides the resonance linewidth, $Q_\text{int}$ is the internal quality factor which was determined in this case by resistive losses in the resonator, and $Q_\text{ext}$ is the external quality factor determined by the capacitance of the resonator to the feedline.
The driving frequency is $\omega$ and the resonance frequency $\omega_r$.
The measured $S_{11}$ is given by
\begin{equation}
	S_{11} = s_{11}+\frac{s_{12}s_{21}}{1-s_{22}S_{11}^{\text{dev}}}S_{11}^{\text{dev}}\ .
\end{equation}
We make the assumption that the reflections $s_{11}, s_{22}$ are very small, such that $s_{22}S_{11}^{\text{dev}} \ll 1$, $s_{11} \ll s_{12}s_{21}S_{11}^{\text{dev}}$, and that $s_{12}s_{21}$ follows an affine function.
This leads to the model which was used to fit the reflection coefficient: $(A+B\omega)S_{11}^{\text{dev}}$.
The fits are shown in figure~\ref{fig:s11_fits}(a).

\begin{figure}
\includegraphics[width=\columnwidth]{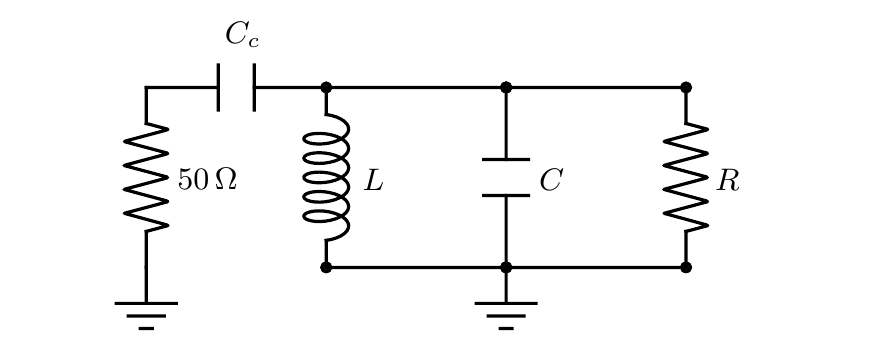}
\caption{%
Lumped-element model of the microwave resonator ($LCR$), with its coupling capacitor $C_c$ and $50\ \Omega$ feedline.
This model is used to quantify the impact of changes in resistivity and dielectric constant on the resonance frequency.
}
\label{fig:circuit}
\end{figure}

\subsection{Metal resistivity}
In the microwave resonator, the majority of the internal losses are due to resistive losses in the metal.
We can therefore extract the internal quality factor from the fitted $S_{11}$ data, and hence track the resistivity $\rho$ of the metal as a function of temperature.
From equation \ref{eq:S11_dev}, we see that the reflection parameter on resonance is determined by the ratio $Q_\text{ext}/Q_\text{tot}$.
The linewidth and depth of the reflection parameter therefore allow us to determine the values of $Q_\text{int}$ and $Q_\text{ext}$.
We make the approximation that $Q_\text{int}$ is entirely determined by resistive losses following $Q_\text{int}\propto 1/\rho$ and that the room temperature resistivity is given by that of bulk gold ($\SI{22}{n\Omega.m}$).

Figure \ref{fig:s11_fits}(a) shows the fitted $S_{11}$ measurements from the first iteration of the surface trap, as a function of temperature.
We observe the transition through the three coupling regimes which was discussed in the main text, and we do not observe the second resonance peak which we have attributed to resonant coupling to the RF electrode.
From the fits, we are able to extract all three quality factors as a function of temperature.
These are shown in figure \ref{fig:s11_fits}(b).
We observe a significant increase in $Q_\text{int}$ at lower temperatures, as expected.
 $Q_\text{tot}$ remains reasonably constant as it is dominated by $Q_\text{ext}$, which does not change significantly as a function of temperature.
This is because $Q_\text{ext}$ is determined by the capacitance of the microwave resonator to the feedline \cite{Teufel2011} which is dominated by geometry, as the dielectric properties of the sapphire substrate vary little with temperature.
 The resulting resistivity as a function of temperature is shown in figure \ref{fig:s11_fits}(c).
Using a polynomial fit to these data points, we conclude that the resistivity is $\SI{13}{n\Omega.m}$ and $\SI{6.5}{n\Omega.m}$ at $\SI{80}{K}$ and $\SI{20}{K}$ respectively.
 Based on this fitted resistivity, the horizontal magnetic field gradient should increase by a factor of 1.45 at $\SI{80}{K}$ and 1.95 at $\SI{20}{K}$.
This is in good agreement with the measured magnetic field gradients displayed in table \ref{tab:gradient_fit}.

\subsection{Frequency shift}

As the temperature decreases from $\SI{295}{K}$ to $\SI{12}{K}$, the resonance frequency increases by $\SI{135}{MHz}$ (figure~\ref{fig:s11_fits}(a)).
This change is at least partially accounted for by the change in metal resistance, quantified above, as well as the change in dielectric constant.
The dielectric constant of sapphire decreases by $\SI{1.5}{\percent}$ between room temperature and liquid helium temperatures \cite{Molla1993}.
A change in dielectric constant modifies the wavelength of the microwaves on the surface trap, which is given by equation \ref{eq:wavelength}.
Since $\epsilon_\text{sap}\gg\epsilon_0$, changing $\epsilon_\text{sap}$ by $\SI{1.5}{\percent}$ modifies the wavelength by approximately a factor $1/\sqrt{0.985}$, or $\SI{0.8}{\percent}$.
This increase in wavelength means that our resonator is observed to be shorter, in units of wavelength, corresponding to an increase in resonance frequency of $\SI{25}{MHz}$.

To quantify other changes in frequency, we use a simplified model of the system.
We use the fact that the input impedance of our $\lambda/4$ resonator is equivalent to that of a parallel $L-C$ circuit \cite{Gely2017} with $L = 4Z_0/\pi\omega_0$ and $C = \pi/(4\omega_0Z_0)$, where $\omega_0$ is the unloaded resonance frequency (close to $\omega_r)$ and $Z_0$ is the characteristic impedance of the waveguide.
From COMSOL simulations, we find $Z_0 = \SI{60}{\Omega}$ where the microwave waveguide is at its widest, which is the value used here.
The model was constructed in QuCAT~\cite{Gely2020}, a Python library which computes resonance frequencies and quality factors of arbitrary circuits.
To the $L-C$ resonator, we added a coupling capacitor $C_c$, a $\SI{50}{\Omega}$ resistor modelling the feedline, and a resistor $R$ in parallel with the $L-C$ network to model internal losses (figure~\ref{fig:circuit}).
The values of $R$ and $C_c$ were chosen to match the internal, external, and total quality factors fitted above.
Note that QuCAT only provides a total quality factor, or rather dissipation rate $\kappa=\omega_r/Q_\text{tot}$, but the internal quality factor was determined by dramatically reducing $C_c$ and extracting $\kappa\approx\omega_r/Q_\text{int}$.
The external quality factor was determined by dramatically increasing $R$ and extracting $\kappa\approx\omega_r/Q_\text{ext}$.
Finally, the unloaded resonator was adjusted to match the resonance frequency measured at room temperature.
We obtained $\omega_0=\SI{3.7}{GHz}$, $R = \SI{1.4}{k\Omega}$, and $C_c = \SI{400}{fF}$, the latter being confirmed through a COMSOL simulation.

This lumped-element model allows us to investigate the impact of the changing dielectric constant and resistance on the resonance frequency.
The impact on the waveguide dielectric constant was already estimated above; however, the dielectric constant also influences the value of $C_c$.
Since $\epsilon_\text{sap}\gg\epsilon_0$, we can assume $C_c\propto \epsilon_\text{sap}$.
Increasing this capacitance by $\SI{1.5}{\percent}$ in our model results in a frequency increase of $\SI{8}{MHz}$.
Increasing the resistance of the $R-L-C$ resonator by a factor 5 (equivalent to a decreasing metal resistivity) results in a frequency increase of $\SI{11}{MHz}$.

Finally, we consider thermal contraction of the sapphire substrate and the copper pillbox to which the substrate is bonded.
From \cite{Baysinger2015}, these two materials should contract by fractional amounts $8\times 10^{-4}$ and $3\times 10^{-3}$ respectively, leading to at most a $\SI{10}{MHz}$ increase in the resonance frequency through a decrease in the length of the resonator.

In total, these changes result in a total frequency increase of $\SI{54}{MHz}$.
Whilst these changes are of the correct order of magnitude, other effects which are not yet understood must be contributing to produce the remainder of the experimentally-observed shift.

\end{document}